\documentclass[amsmath,amssymb, preprint, Lucida Grande, 12pt, letter paper]{revtex4-1}
\usepackage{graphicx,subfigure, float, tabularx, dcolumn, bm, epsfig, amsmath, multirow, array}
\usepackage[compact]{titlesec}
\makeatletter

\newcommand{\Rmnum}[1]{\expandafter\@slowromancap\romannumeral #1@}
\newcommand{\dotr}{\mbox{$\boldsymbol{\cdot}$}}
\usepackage{color}
\usepackage{soul}
\sethlcolor{yellow}
\makeatother
\begin{document}
\title{First-Principles Calculation of the Bulk Photovoltaic Effect in KNbO$_{3}$ and (K,Ba)(Ni,Nb)O$_{3-\delta}$}
\author{Fenggong Wang}
\email[]{fenggong@sas.upenn.edu}
\author{Andrew M. Rappe}
\email[]{rappe@sas.upenn.edu}
\affiliation{The Makineni Theoretical Laboratories, Department of Chemistry, University
of Pennsylvania, Philadelphia, PA 19104--6323}
\date{\today}
\begin{abstract}
The connection between noncentrosymmetric materials' structure, electronic structure, and bulk photovoltaic performance remains not well understood. In particular, it is still unclear which photovoltaic (PV) mechanism(s) are relevant for the recently demonstrated visible-light ferroelectric photovoltaic (K,Ba)(Ni,Nb)O$_{3-\delta}$. In this paper, we study the bulk photovoltaic effect (BPVE) of (K,Ba)(Ni,Nb)O$_{3-\delta}$ and KNbO$_{3}$ by calculating the shift current from first principles. The effects of structural phase, lattice distortion, oxygen vacancies, cation arrangement, composition, and strain on BPVE are systematically studied. We find that (K,Ba)(Ni,Nb)O$_{3-\delta}$ has a comparable BPVE with that of the broadly explored BiFeO$_{3}$, but for a much lower photon energy. In particular, the Glass coefficient of (K,Ba)(Ni,Nb)O$_{5}$ in a simply layered structure can be as large as 12 times that of BiFeO$_{3}$. Furthermore, the nature of the wavefunctions dictates the eventual shift current yield, which can be significantly affected and engineered by changing the O vacancy location, cation arrangement, and strain. This is not only helpful for understanding other PV mechanisms that relate to the motion of the photocurrent carriers, but also provides guidelines for the design and optimization of PV converters.
\end{abstract}
\maketitle
\section{INTRODUCTION}
As the world power consumption and carbon emissions continue to increase, solar energy has drawn even more attention because it is clean, abundant, and sustainable, and thus is widely seen as a long-term substitute for traditional fossil fuels~\cite{Maeda06p295}.
Efficient solar energy conversion relies primarily on semiconducting photoabsorbers with a low band gap ($E_{g}$), allowing absorption of most of the solar light reaching the earth.
With light absorption, electrons are excited to the material's conduction band (CB) for electricity generation or catalysis~\cite{Choi09p63, Kudo09p253, Wang12p476}.
However, the photo-excited electrons may also recombine radiatively or nonradiatively  with the created holes, reducing the power conversion efficiency.
In conventional solar cells, this recombination rate is minimized by an externally engineered asymmetry, i.e., electrons and holes are separated by the electric field in a $p-n$ junction or other interface.
This not only complicates the device fabrication, but also imposes the Shockley$-$Queisser limit on the power conversion efficiency of this type of device~\cite{Shockley61p510}.
Ferroelectric (FE) solar converters, lacking inversion symmetry due to intrinsic spontaneous polarization, can separate photo-excited charges by the depolarization field or by the bulk photovoltaic effect (BPVE)~\cite{Glass74p233, Kraut79p1548, Chynoweth56p705}.
In the BPVE, a spontaneous direct short-circuit photocurrent is generated when electrons are continuously excited to a quasiparticle coherent state that has an intrinsic momentum, avoiding the need for an interface to separate charge.
In particular, the BPVE is able to generate an above-band-gap photovoltage~\cite{Ji10p1763}, potentially enabling a higher power conversion efficiency than the Shockley$-$Queisser limit. 

However, most conventional ferroelectric materials have wide band gaps [$E_{g}>$2.7 eV for BiFeO$_{3}$, $E_{g}>$3.5 eV for Pb(Zr$_{1/2}$Ti$_{1/2}$)O$_{3}$], limiting their ability to absorb the visible light that makes up the biggest fraction of the solar irradiance.
Thus, an enormous amount of efforts has been focused on the design and optimization of FE materials in order to reach a lower band gap~\cite{Bennett08p17409, Gou11p205115, Li13p823, Yang12p15953, Berger12p165211, Zhang10p012905, Choi12p689, Xu10p192901, Kan11p1765, Wang12p8901, Nechache11p202902, Takagi11p115130, Jiang14p075153, Wang14p152903, Wang14p235105, Wang12p476, Wang14p152903,Qi11p224108}.
 Among them, the study and improvement of ferroelectric oxides are particularly important, as these materials can be integrated into the conventional electronics~\cite{Nonaka99p1143, Yang10p143, Qin08p122904, Alexe11p256, Zhang13p2109, Zhang13p1265, You14p1674, Snaith13p3623, Stoumpos13p9019}.
In particular, recently a ferroelectric perovskite (K,Ba)(Ni,Nb)O$_{3-\delta}$ (KBNNO) has been successfully synthesized and demonstrated to have a near-optimal band gap (1.39 eV), exhibiting good photovoltaic (PV) performance~\cite{Grinberg13p509}.
While this has substantially advanced the area of ferroelectric photovoltaics, there remain open questions. For example, what is the underlying PV mechanism in this material?
Furthermore, this material is also able to exhibit BPVE, as previous time dependent perturbation theory analysis has shown that BPVE, in principle, can arise in any polar material through the ``shift current'' mechanism~\cite{Baltz81p5590, Young12p116601}.
Actually this becomes even more fascinating when taking into account that the parent KNbO$_{3}$ is an interesting nonlinear optical (NLO) material with high NLO coefficients~\cite{Fluckiger87p406, Donafeng97p1399}.
In this paper, we study the BPVE and its correlation to structural and electronic properties in KBNNO and KNbO$_{3}$ from first principles.
The connection between the photocurrent and electronic structure elucidated here is not only useful for understanding other PV mechanisms in KBNNO in the sense that all these PV mechanisms relate to the light absorption and the motion of the photo-excited carriers and thus to the electronic properties, but also can be generalized to other similar materials that have great tunability of orbital character near the band gap. 
\section{COMPUTATIONAL DETAILS}
Density functional theory (DFT) calculations were performed with the local density approximation (LDA) functional, as implemented in the QUANTUM-ESPRESSO code~\cite{Giannozzi09p395502, Kohn65pA1133, Perdew81p5048}.
Norm-conserving, optimized nonlocal pseudopotentials were used to represent all elements~\cite{Rappe90p1227}.
The DFT+$U$ method was used to improve the description of $d$-orbital electrons by better accounting for the correlation effect, with Hubbard $U$ parameterized by the linear-response approach~\cite{Cococcioni05p035105}.
The calculated Hubbard $U$ values are 3.6 eV for Nb in KNbO$_{3}$, and 3.97 eV and 9.90 eV for Nb and Ni in KBNNO.
The shift current was calculated with a previously developed first-principles approach based on time-dependent perturbation theory, which yields good agreement with experiment for the prototypical ferroelectric oxides, such as BiFeO$_{3}$~\cite{Young12p116601, Young12p236601}.
To calculate the shift current, a self-consistent calculation was first done by the LDA+$U$ approach for the structures fully relaxed by LDA, followed by non-self-consistent calculations with much finer $k$ grids.
We used the LDA relaxed structures for shift current calculations, as LDA was shown to describe well the KNbO$_{3}$ structural properties, with only 0.3\% underestimation of the tetragonality $c/a$~\cite{Wang14p235105}.
The Monkhorst-Pack $k$-mesh method~\cite{Monkhorst76p5188} was used to sample the Brillouin zone.
To converge the shift current, the $k$ grid must be sufficiently dense, e.g., a 40$\times40\times$40 $k$ grid was used for a typical $AB$O$_{3}$ unit cell.
In order to overcome the self-interaction error of the standard DFT method, we also used the HSE06 hybrid functional to calculate the band gaps of some compositions~\cite{Heyd03p8207}.
The HSE06 hybrid functional improves the band gap description by including a proportion of the exact exchange interaction, while the correlation part remains the same as in the standard DFT method; it only serves as a corroboration of the band gap here. We also adopted the GGA+$U$ method to calculate the shift current of KNbO$_{3}$ and some KBNNO structures, and confirmed that the shift current spectral features are essentially similar to those obtained by the LDA+$U$ method (the change of the shift current magnitude is tiny), except for slightly larger onset photon energies in the GGA+$U$ case.
\section{RESULTS AND DISCUSSIONS}
\subsection{The BPVE of KNbO$_{3}$}
\begin{figure}[b]
\centering
\subfigure{\includegraphics[width=0.49\textwidth]{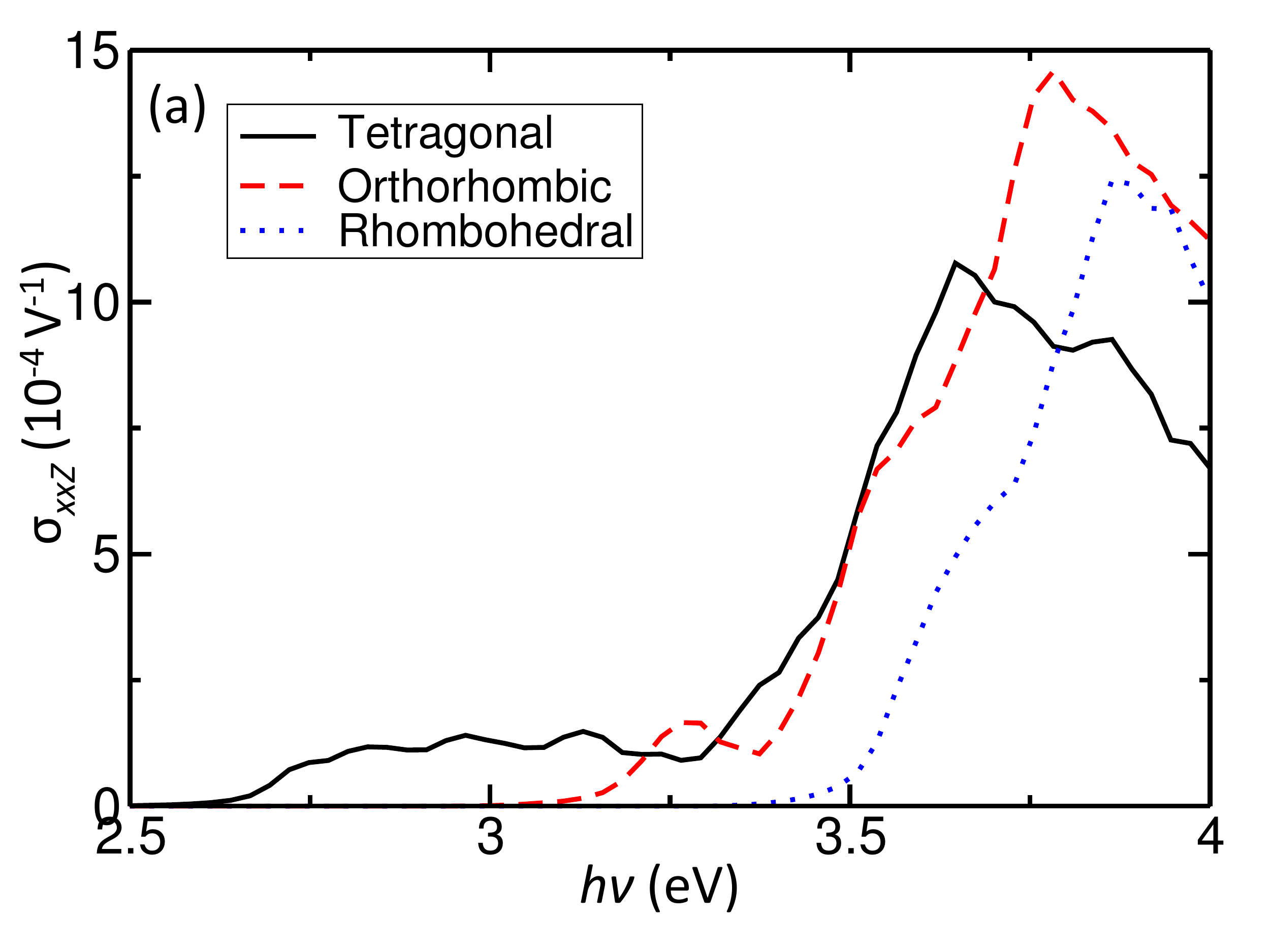}}
\subfigure{\includegraphics[width=0.49\textwidth]{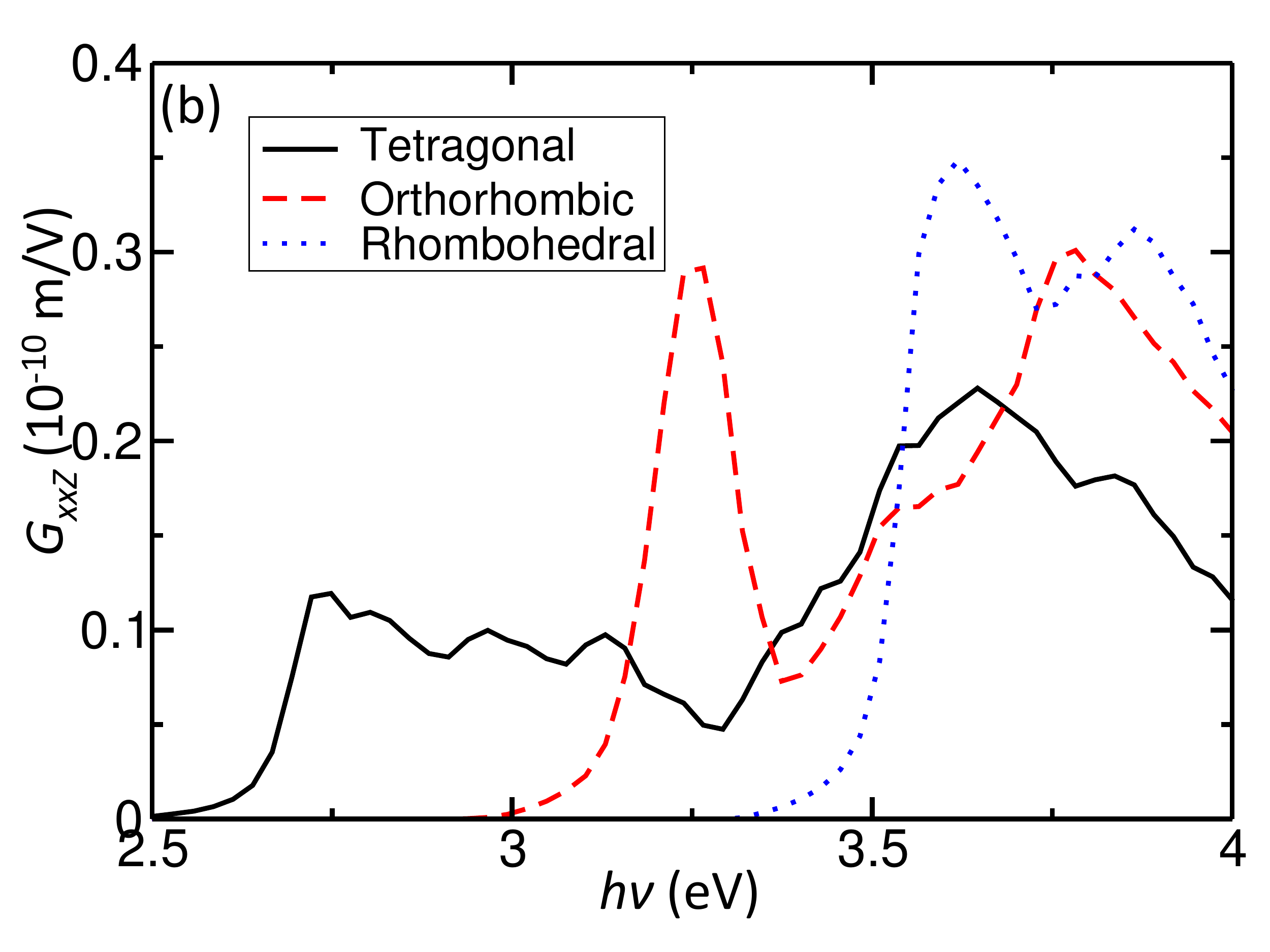}}
\caption{(Color online) The (a) shift current susceptibility and (b) Glass coefficient of various phases of KNbO$_{3}$.
\label{SC-KNO}}
\end{figure}
Shift current, as a dominant mechanism for the BPVE, is a second-order nonlinear optical effect with the photocurrent quadratic in the electric field ($J_q = \sigma_{rsq} E_r E_s$)~\cite{Young12p116601}. 
The Glass coefficient
%
\begin{align}
G_{rrq}=&\frac{\sigma_{rrq}(\omega)}{\alpha_{rr}(\omega)}
\end{align}
%
describes the current response in a thick sample and includes the light attenuation effect due to the absorption coefficient $\alpha_{rr}(\omega)$.

We start by calculating the shift current of the parent material KNbO$_{3}$.
KNbO$_{3}$ is a typical $AB$O$_{3}$ perovskite ferroelectric oxide that occurs in various different phases.
At temperatures above 691 K, KNbO$_{3}$ is in a paraelectric cubic phase with space group $Pm\bar{3}m$.
As the temperature decreases below 691 K, it first undergoes a phase transition into a tetragonal phase (space group $P4mm$), and then into an orthorhombic phase at 498 K (space group $Amm2$) and an even more distorted rhombohedral phase with space group $R3m$ at 263 K.
Concurrently with the structural transitions from cubic to tetragonal, orthorhombic, and rhombohedral phases, the NbO$_{6}$ octahedra become more severely distorted through octahedral tilting and rotation, and the Nb ions move away from the center of the O$_{6}$ cages.
As a typical $AB$O$_{3}$ perovskite semiconductor/insulator, the valence band maximum (VBM) of KNbO$_{3}$ arises mainly from the O 2$p$ orbitals, while the conduction band minimum (CBM) is predominantly composed of Nb 4$d$ orbitals~\cite{Cabuk07p100, Okoye03p5945}.
Also, the band gap of KNbO$_{3}$ increases significantly when the structure changes from cubic to the more distorted rhombohedral phase, because of the correlation between the NbO$_{6}$ octahedral distortion and Nb ions off-center displacements and the electronic structure~\cite{Wang14p152903}.
Correspondingly, experiment finds a wide range of band gap values, from 3.3 eV for the cubic phase~\cite{Wiesendanger74p203} to 4.4 eV for the tetragonal phase~\cite{Calendiri81p1179}.
Our previous HSE06 calculation also shows that the band gap of the rhombohedral KNbO$_{3}$ is $\approx$0.6 eV larger than that of its tetragonal counterpart (Table~\ref{tab:HSE})~\cite{Wang14p152903}.
\begin{table}[t]
\caption{The HSE06 band gap of various KNbO$_{3}$ phases (cubic, tetragonal, orthorhombic, and rhombohedral) and two different cation arrangements ($1\times1\times2$ and rocksalt) of the (K,Ba)(Ni,Nb)O$_{5}$ (KBNNO) solid solutions. KBNNO has much smaller band gaps than KNbO$_{3}$. All structures are fully relaxed with LDA.}
\centering
\begin{tabular}{lccccccc}
\hline\hline
\multirow{2}{*}{System}&\multicolumn{3}{c}{KNbO$_{3}$}&&\multicolumn{2}{c}{(K,Ba)(Ni,Nb)O$_{5}$}\\
\cline{2-5}\cline{7-8}
&Cubic&Tetragonal &Orthorhombic &Rhombohedral &&$1\times1\times2$&rocksalt\\
\hline
$E^{\rm HSE}_{g}$&2.65&2.71&3.10&3.33&&1.28&2.16\\
\hline\hline
\end{tabular}
\label{tab:HSE}
\end{table}

\begin{figure}[b]
\centering
\includegraphics[width=\textwidth]{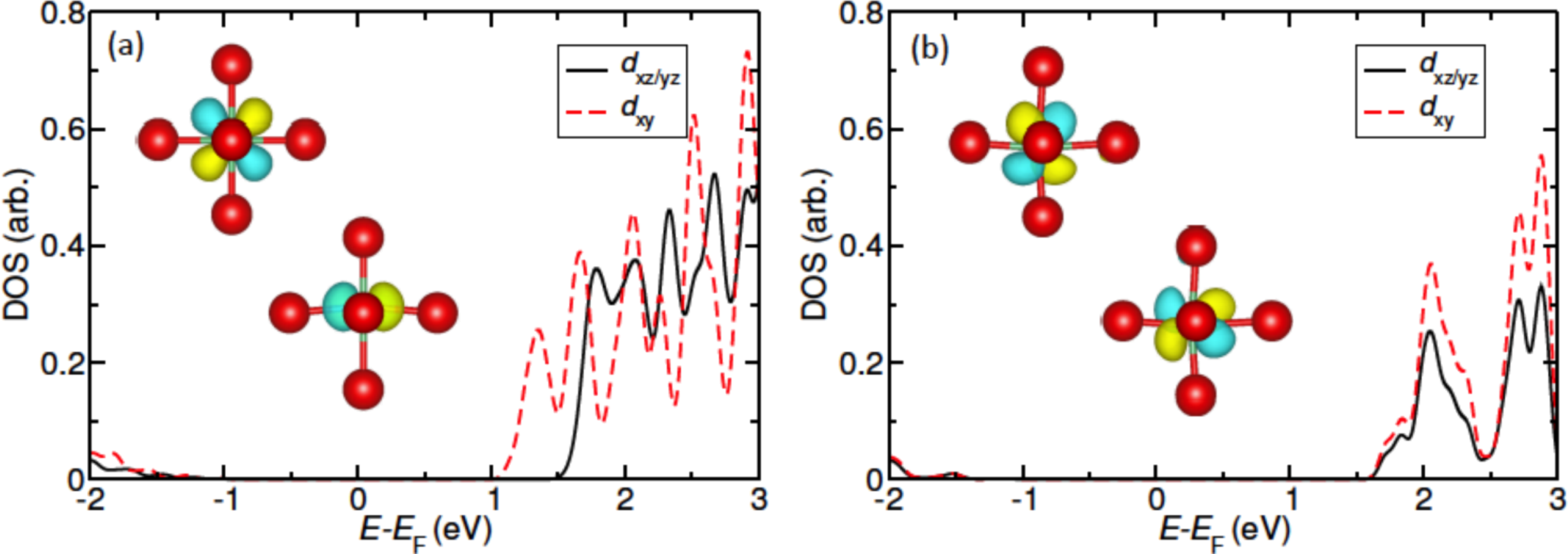}
\caption{(Color online) The projected densities of states (PDOSs) onto the Nb $d$ orbitals of the (a) tetragonal and (b) rhombohedral KNbO$_{3}$. The inset shows the corresponding real-space wavefunction distribution around the NbO$_{6}$ octahedra for the conduction band minimum at the $\Gamma$ point in the Brillouin zone. Upper: view along $z$; Lower: view along $x$. 
\label{KNO-PDOS}}
\end{figure}
Figure~\ref{SC-KNO} shows the calculated shift current susceptibilities and Glass coefficients of KNbO$_{3}$ in its ferroelectric tetragonal, orthorhombic, and rhombohedral phases.
For convenience, we only shows the largest tensor element \emph{xxZ}, where the upper case letter represents the shift current direction.
Clearly, the magnitude of the maximum shift current susceptibility for the room-temperature orthorhombic phase with respect to the photon energies between $E_{g}$ and $E_{g}$+1.0 eV ($\approx$15$\times$10$^{-4}$ V$^{-1}$) is more than twice that of the more broadly studied BiFeO$_{3}$ ($\approx$6$\times$10$^{-4}$ V$^{-1}$), although their band-edge shift current responses are on a par with each other ($\approx$1$\times$10$^{-4}$ V$^{-1}$).
This indicates that KNbO$_{3}$ is more promising than BiFeO$_{3}$ for photovoltaic applications with high photon energies (ultraviolet light), but it is not as good as BiFeO$_{3}$ for light at the visible-UV edge (3.0-3.2 eV), since BiFeO$_{3}$ has a much lower band gap (2.7 eV). 
Following the band gap change as structural phase is changed, the onset photon energies of both the shift current susceptibility and the Glass coefficient are the lowest in the tetragonal phase, followed by that in the orthorhombic and rhombohedral phases, suggesting that the greater the lattice distortions, the higher the energy required to trigger a shift current response.
It should be pointed out that the symmetric cubic phase is not able to exhibit shift current, although it has a lower band gap.
Furthermore, for the same photon energies that are $\approx$1.2 eV above the band gap of the tetragonal phase ($E_{g}^{0}$+1.2 eV, $E_{g}^{0}$ is the LDA+$U$ band gap of tetragonal KNbO$_{3}$), all three phases exhibit comparable magnitude of both the shift current susceptibility and the Glass coefficient, with the response in the tetragonal phase slightly smaller than those in the other two phases. 

It is noteworthy that the band-edge (different $E_{g}$, and thus different photon energies) shift current response and Glass coefficient of the tetragonal phase are much smaller than those of the other two phases.
Figure~\ref{KNO-PDOS} shows the projected density of states (PDOS) onto the Nb $d$ orbitals and the real-space wavefunction isosurfaces of the CBM at $\Gamma$ $k$ point for the tetragonal and rhombohedral phases.
Even though both the tetragonal and rhombohedral phases possess CBM composed of Nb $d$ orbitals, they are subtly very different.
The CBM of the tetragonal phase arises mainly from the Nb $d_{xy}$ orbital, with a 0.5 eV energy splitting between the $d_{xy}$ and $d_{zx/zy}$ orbitals.
However, in the rhombohedral phase, there is almost no splitting between the $d_{xy}$ and $d_{zx/zy}$ orbitals, leading to a mixing of both $d_{xy}$ and $d_{zx/zy}$ orbitals in the CBM.
This occurs because in the rhombohedral phase the Nb atom moves away from the O$_{6}$ cage center along all three Cartesian directions, reducing the difference in the Nb-O bonding strength in the different directions.
In contrast to the $d_{xy}$ orbital, the $d_{zx/zy}$ orbitals extend the wavefunction along the shift current direction ($z$), facilitating the motion of the shift current carriers.
Consequently, the shift current susceptibility and Glass coefficient are larger in the rhombohedral phase than in the tetragonal phase.
\subsection{Effect of O vacancy and cation arrangement on shift current in KBNNO}
%
\begin{figure}[t]
\centering
\includegraphics[width=0.7\textwidth]{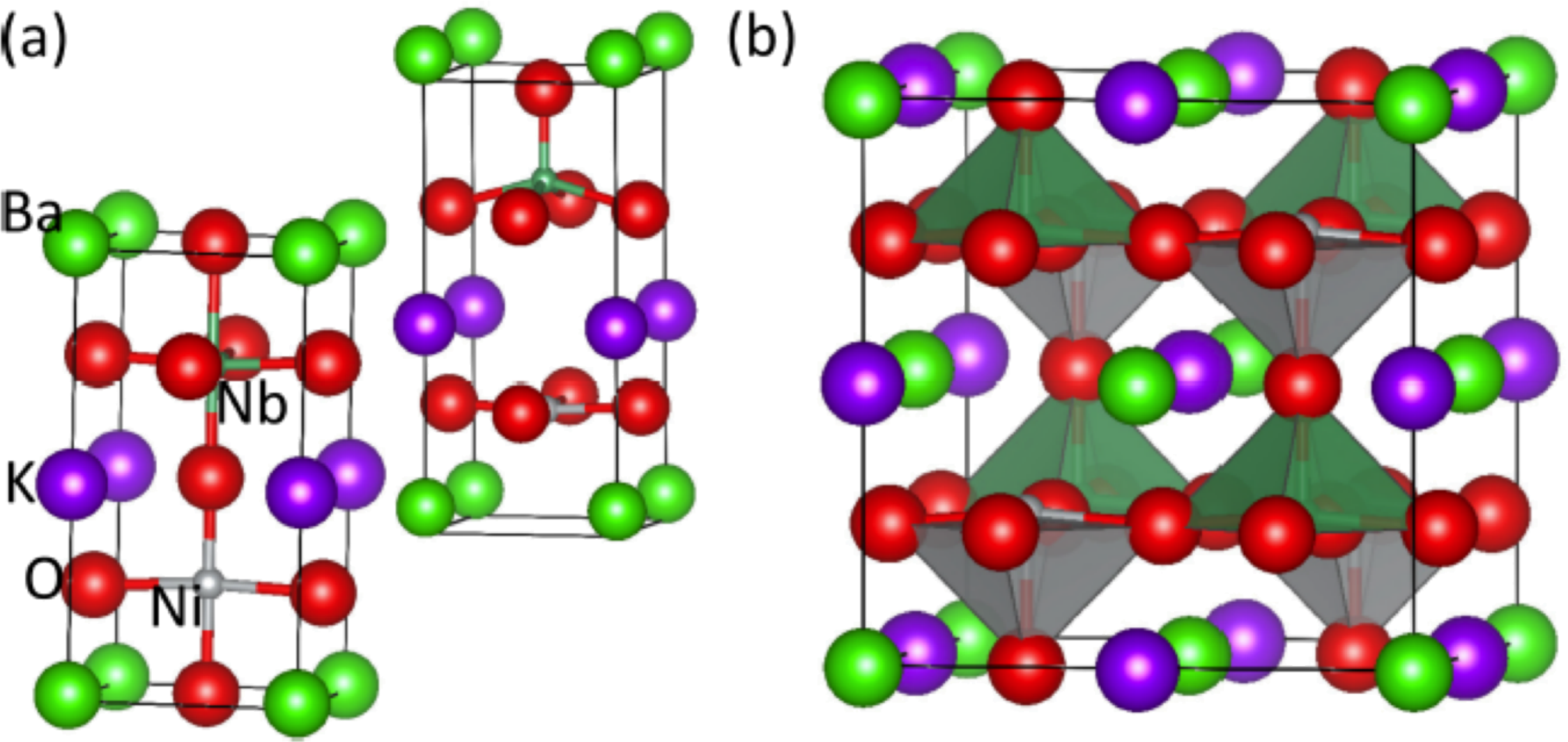}
\caption{(Color online) The atomic structure representation of the (K,Ba)(Ni,Nb)O$_{5}$ solid solution with (a) layered and (b) rocksalt cation arrangements. Atoms are drawn with accepted ionic radii. In (a), both the apical and equatorial O vacancies are shown. 
\label{2KNO-Struc}}
\end{figure}
\begin{table}[t]
\caption{The relative total energies (meV) of different O vacancy sites in tetragonal KBNNO. For the $1\times1\times N$ layered structure, the top apical site is in the K layer and the bottom apical site is in the Ba layer.}
\centering
\begin{tabular}{lccc}
\hline\hline
 Supercells&Top-apical&Bottom-apical&Equatorial\\
\hline
$1\times1\times2$&+115&+143&0\\ 
Rocksalt&0&0&0\\
$1\times1\times3$&+94&+124&0\\
$1\times1\times4$&+77&+103&0\\
\hline\hline
\end{tabular}
\label{tab:o-vacancy}
\end{table}
In KBNNO, some of the original Nb$^{5+}$ ions are substituted by Ni$^{2+}$ ions (Ni$_{\rm Nb}^{'''}$), with the charge compensated by the combination of $A$-site substitution of Ba$^{2+}$ for K$^{+}$ ions (Ba$_{\rm K}^{\dotr}$) and O vacancies ($V_{\rm O}^{\dotr\dotr}$).
Generally, the O vacancies prefer to form adjacent to the Ni$^{2+}$ dopant, because the possible O vacancy as a donor is attracted by the Ni$_{\rm Nb}^{'''}$ acceptor.
However, there are still three inequivalent O vacancy sites adjacent to Ni: the top apical site (with the Ni$_{\rm Nb}^{'''}$-$V_{\rm O}^{\dotr\dotr}$ local polarization parallel to the overall polarization), the equatorial site, and the bottom apical site (with the Ni$_{\rm Nb}^{'''}$-$V_{\rm O}^{\dotr\dotr}$ local polarization antiparallel to the overall polarization).
We study extensively the stabilities of different O vacancy sites by using supercells with different compositions or cation arrangements, including both layered and rocksalt $B$-cation structures (Fig.~\ref{2KNO-Struc}).
As shown in Table~\ref{tab:o-vacancy}, in the $1\times1\times N$ layered supercells, the equatorial O vacancy site is much more stable than both the top and bottom apical sites.
This is because when the O vacancy is located at the equatorial site, a network of -Ni-V$_{\rm O}$-Ni- forms.
The attractive Coulomb interaction between the donor $V_{\rm O}^{\dotr\dotr}$ and acceptor Ni$_{\rm Nb}^{'''}$ decreases the total energy, resulting in a more nearly uniform charge distribution than the other two cases.
Also, the top apical site is slightly more favorable than the bottom apical site because of the repulsive interaction between the $V_{\rm O}^{\dotr\dotr}$ and Ba$_{\rm K}^{\dotr}$ donors.
However, in the rocksalt arrangement, there is no preferred O vacancy site, as the charge environment is nearly isotropic for different orientations of the Ni-V$_{\rm O}$ complex.
Even though the Ni$_{\rm Nb}^{'''}$-$V_{\rm O}^{\dotr\dotr}$ local polarization can be parallel or antiparallel to the overall polarization, it has minor effect on the preference of the O vacancy site.  
Basically, the rocksalt arrangement is about 100 meV/atom less stable than the $1\times1\times2$ layered arrangement with equatorial O vacancies, but more stable than that with apical O vacancies.
Overall, we see a clear effect of the Coulomb interaction and charge compensation mechanism on determining the favorable O vacancy site.
\begin{figure}[t]
\centering
\subfigure{\includegraphics[width=0.49\textwidth]{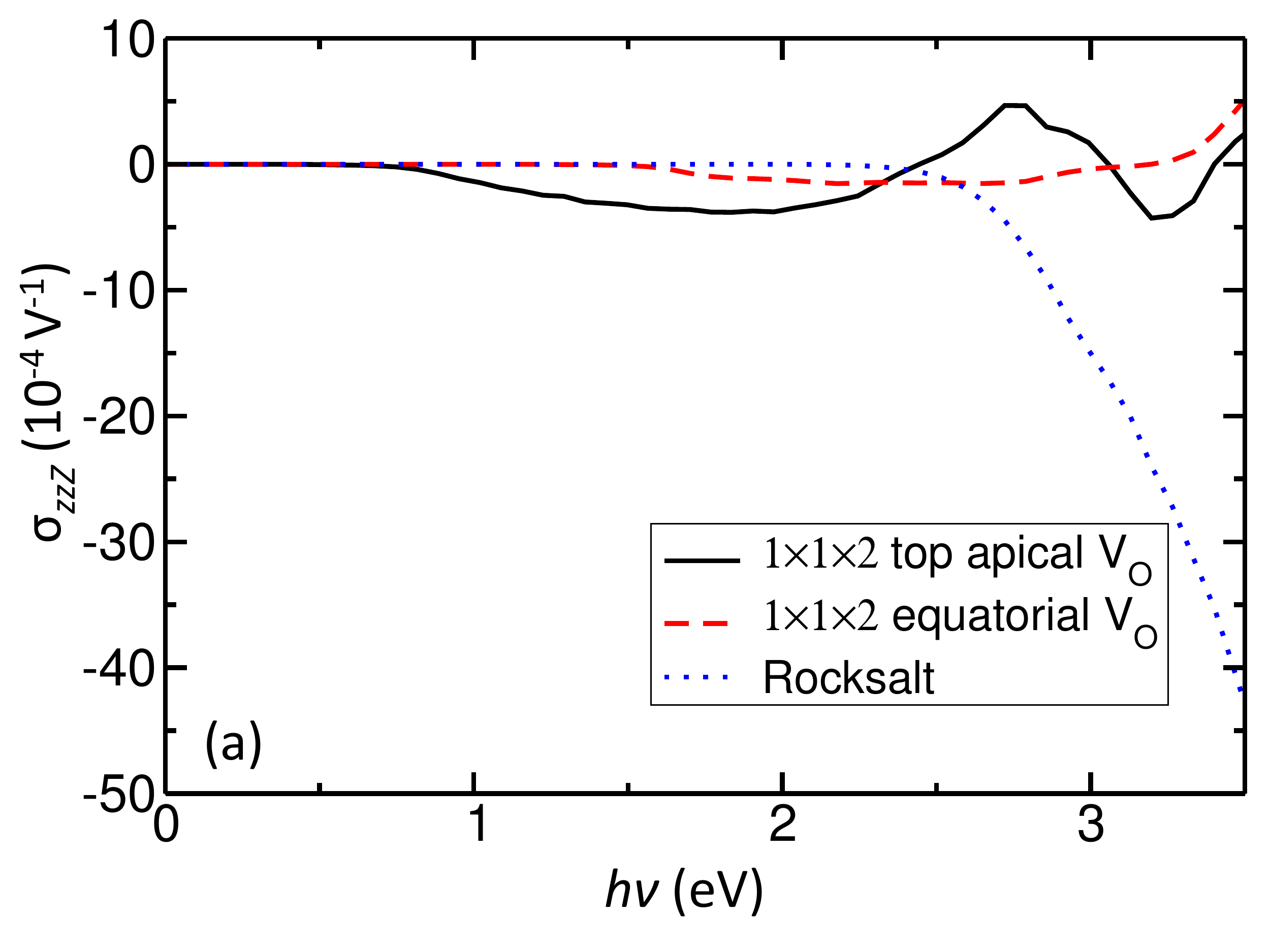}}
\subfigure{\includegraphics[width=0.49\textwidth]{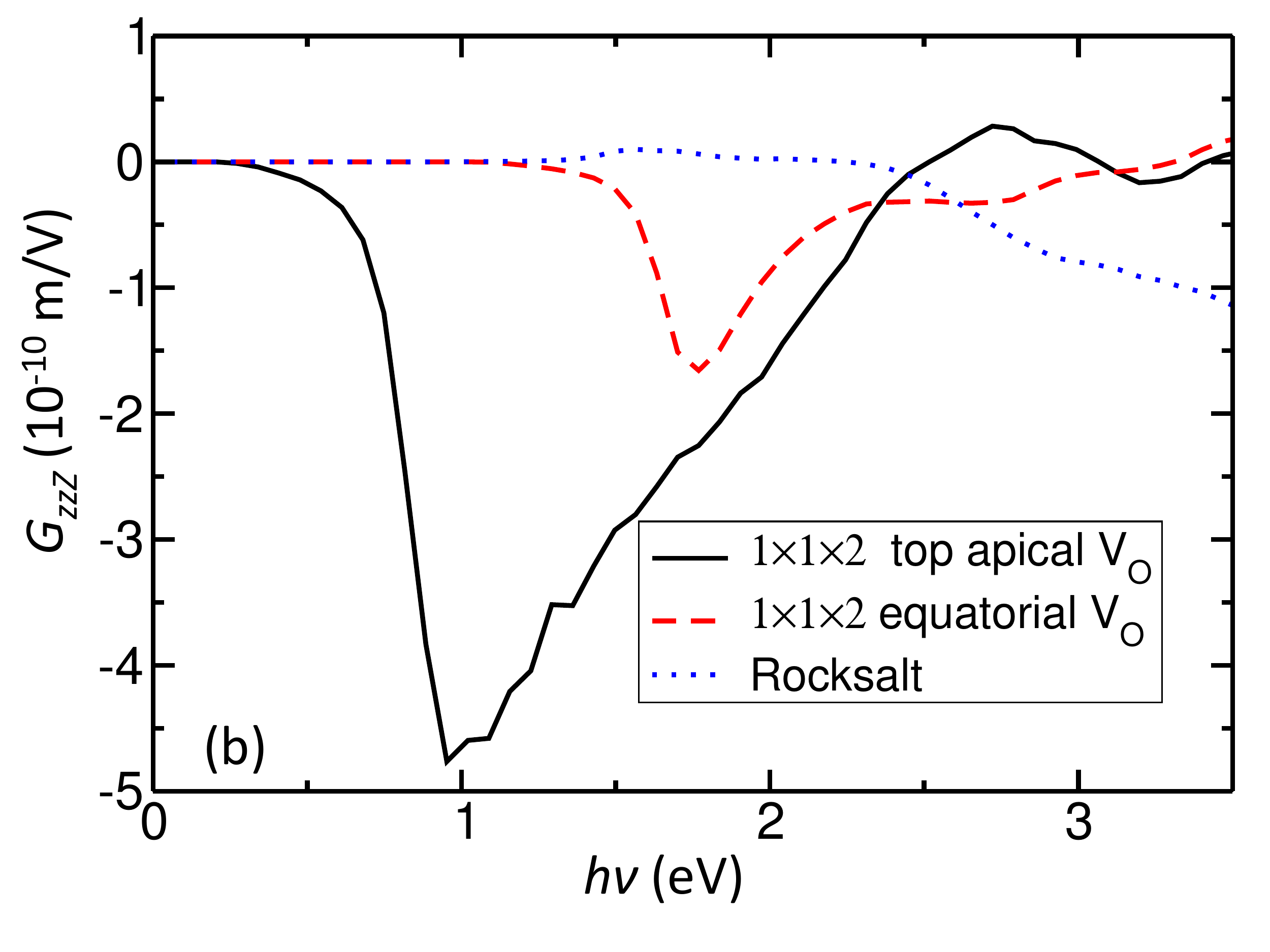}}
\subfigure{\includegraphics[width=0.49\textwidth]{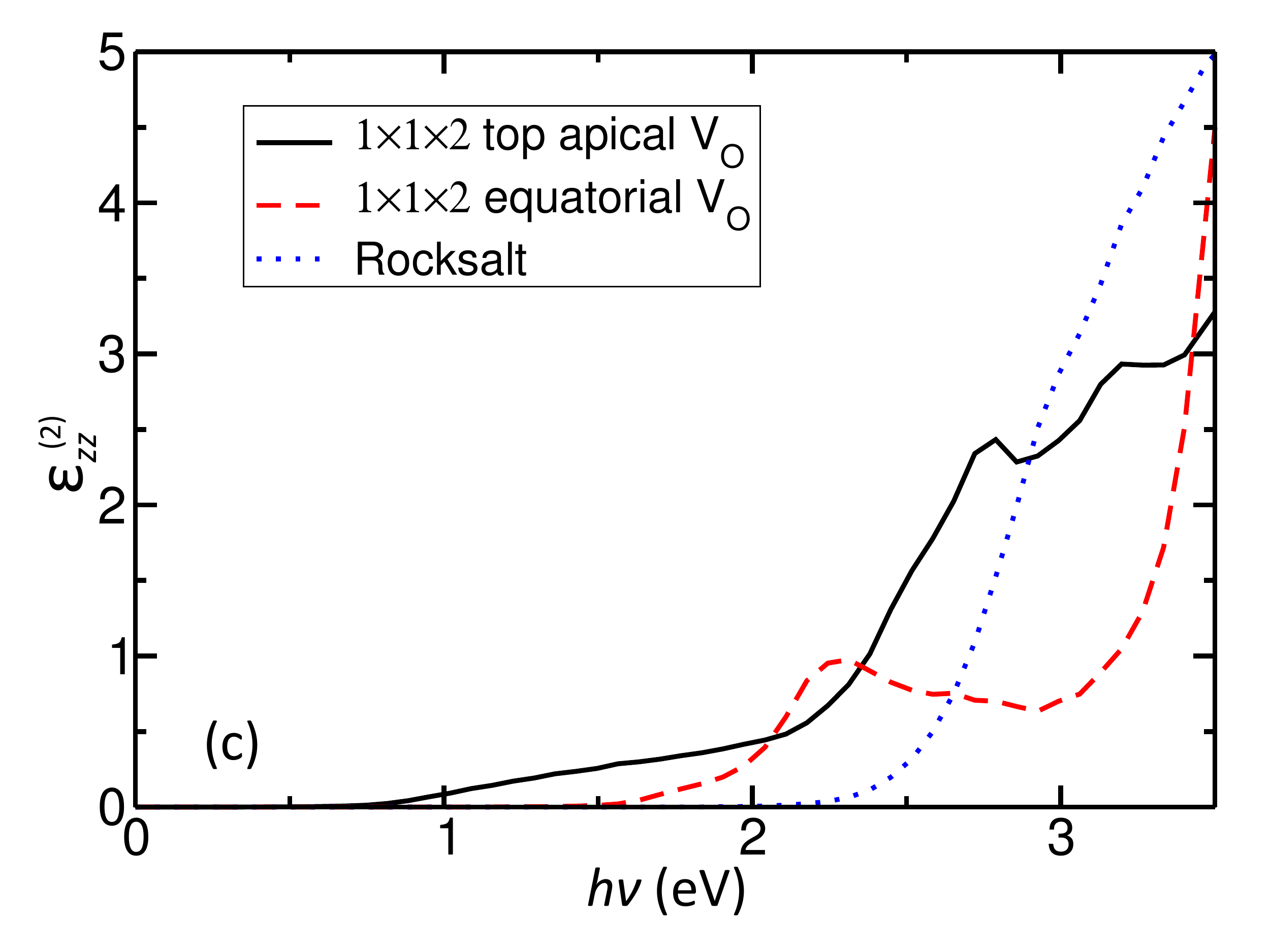}}
\subfigure{\includegraphics[width=0.49\textwidth]{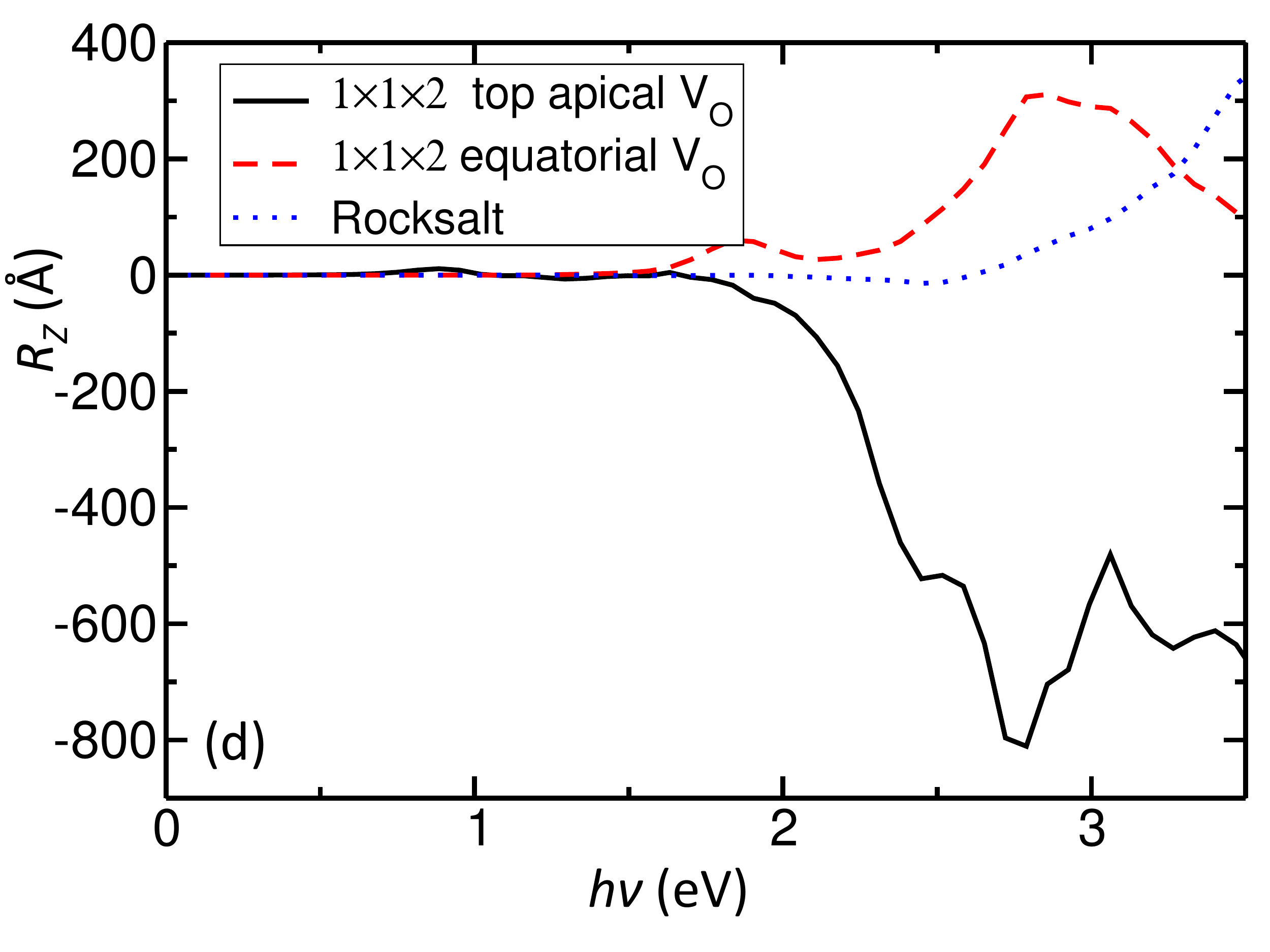}}
\caption{(Color online) The largest tensor element of (a) shift current susceptibility ($\sigma_{zzZ}$) and (b) Glass coefficient ($G_{zzZ}$), and the (c) imaginary dielectric constant ($\epsilon^{(2)}_{zz}$), and (d) shift vector ($R_{Z}$) of the (K,Ba)(Ni,Nb)O$_{5}$ solid solution with $1\times1\times2$ layered and rocksalt cation arrangements. For the $1\times1\times2$ layered arrangement, both the apical and equatorial O vacancy sites are included, while for the rocksalt arrangement the vacancy is at the apical site.
\label{SC-2unit}}
\end{figure}

\begin{figure}[t]
\centering
\includegraphics[width=\textwidth]{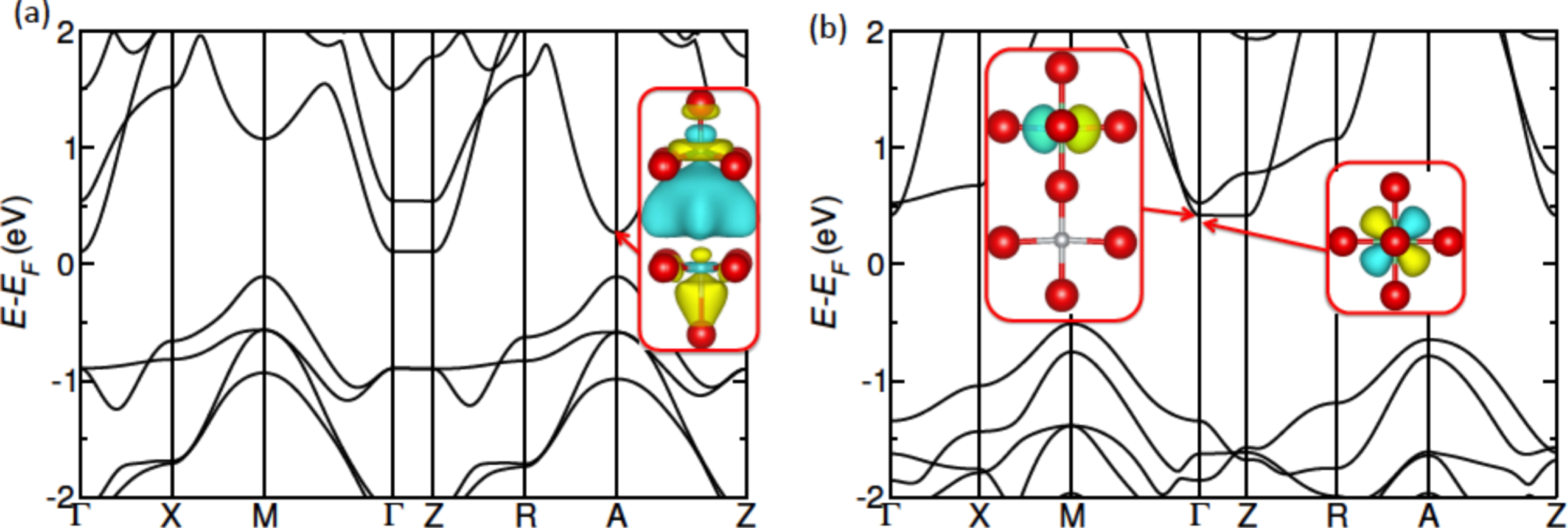}
\caption{(Color online) The band structure of the $1\times1\times2$ layered (K,Ba)(Ni,Nb)O$_{5}$ solid solution (a) with apical and (b) equatorial O vacancies. The inset shows the real-space wavefuction distribution for the corresponding conduction electronic states at $A$(0.5, 0.5, 0.5) $k$ point in (a) and $\Gamma$ $k$ point in (b). The wavefunction plot in (a) is in view along $x$, while the left and right insets in (b) are in views along $x$ and $z$, respectively. The K and Ba ions are not shown in the wavefunction plot.
\label{Fig3-4}}
\end{figure}

To study the effect of different oxygen vacancy sites and cation arrangements on shift current, we calculate the shift current of the $1\times1\times2$ layered supercells both with equatorial and apical O vacancies, as well as the rocksalt $B$-cation arrangement.
Although the layered structures with a high concentration of vacancies may be experimentally difficult to synthesize under normal conditions, they serve as good examples to elucidate the stability of O vacancies and their effects on shift current.
Figure~\ref{SC-2unit} shows the calculated shift current susceptibilities and Glass coefficients of these three different structures of the (K,Ba)(Ni,Nb)O$_{5}$ solid solution.
Several features are clear from the comparison of these results.
First, the Glass coefficient of the layered arrangement with apical O vacancies (5$\times$10$^{-10}$ m/V) is approximately 12 times larger than that of the prototypical ferroelectric photovoltaic BiFeO$_{3}$ (0.4$\times$10$^{-10}$ m/V) for the photon energies between its $E_{g}$ and $E_{g}$+1.0 eV, but the required photon energies in KBNNO are much lower, at the LDA+$U$ level of 1.0 eV. The low band gap of this KBNNO solid solution is further corroborated by its HSE06 band gap of 1.28 eV (Table~\ref{tab:HSE}).
Therefore, this cation arrangement would be a great bulk photovoltaic in a thick sample if the vacancy locations can be controlled.
Compared to the Glass coefficient spectrum, there is not any peak in the shift current susceptibility spectrum or in the shift vector spectrum for photon energies near 1 eV.
The Glass coefficient is large because the absorption coefficient at these energies is extremely small.
These electronic transitions are mainly from the O 2$p$ orbital dominated valence band (VB) to the CB that is composed of Nb 4$d_{z^{2}}$ and Ni 3$d_{z^{2}}$ orbitals as well as O 2$p_{z}$. The transitions near the $A$(0.5, 0.5, 0.5) point of the Brillouin zone are not localized at the sublattice around Ni, but broadly distributed over the whole lattice [Fig.~\ref{Fig3-4}(a)].
%

%
Second, both the shift current susceptibility and Glass coefficient of the $1\times1\times2$ layered KBNNO with apical O vacancies are much larger than those of layered KBNNO with equatorial O vacancies for almost the whole spectral range.
This is because both the imaginary dielectric constant and the shift vector of the layered  apical vacancy structure are larger than those of the layered equatorial vacancy one.
Figure~\ref{Fig3-4} shows the band structure of the $1\times1\times2$ layered KBNNO solid solutions with equatorial and apical O vacancies.
Unlike the apical case, where the band-edge transitions are near the $A$ point, in the equatorial structure they mainly occur near the $\Gamma$ point.
In the latter case, the CBM has a major contribution from the Nb 4$d_{xy}$ orbital [Fig.~\ref{Fig3-4}(b)].
Unlike the $d_{z^{2}}$ orbital, the $d_{xy}$ orbital is distributed within the $xy$ plane that is perpendicular to the shift current direction ($z$), leading to a much smaller onset Glass coefficient for the equatorial case. 
When the equatorial O vacancies are organized along the $x$ direction, a chain of -Ni-V$_{\rm O}$-Ni- formed.
Since the remaining NiO$_{4}$ complex prefers a square planar symmetry in the perpendicular $yz$ plane, the overall lattice asymmetry along the $z$ direction is significantly reduced.
On the other hand, removing the O atom at apical site leads to a larger $c$ lattice constant and an overall enhancement of the lattice asymmetry along the $z$ direction.
The combined effect of the different lattice asymmetries and the orbital character change dictates the change of the shift current magnitude.

\begin{figure}[t]
\centering
\includegraphics[width=\textwidth]{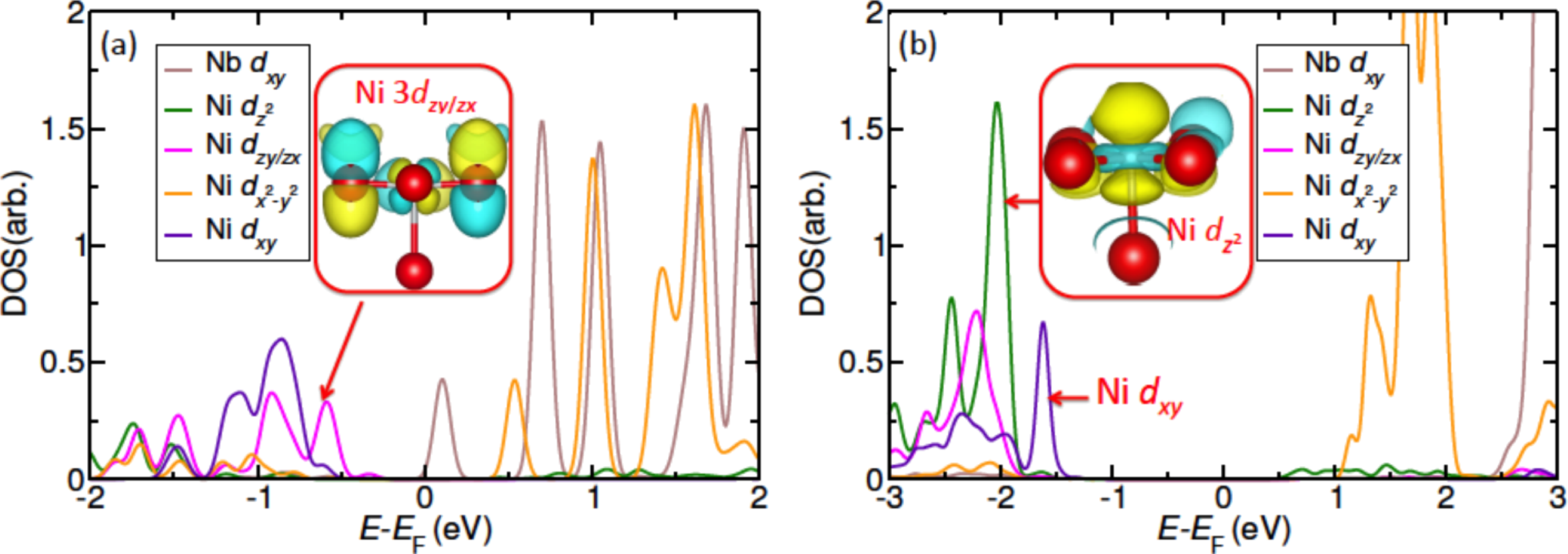}
\caption{(Color online) The projected densities of states (PDOSs) of the (K,Ba)(Ni,Nb)O$_{5}$ solid solution with $1\times1\times2$ layered and rocksalt cation arrangements. The inset shows the real-space wavefunction distribution (view along $x$) for the corresponding electronic states as indicated by the red arrow. In both cases, the O vacancy is at the top apical site. In (a), the equatorial O atom nearest to the viewer is hidden in order to show the orbital character of the Ni 3$d_{zy/zx}$ orbitals.
\label{PDOS-112-RS}}
\end{figure}
%
Also, comparison of the layered and rocksalt cation arrangements, both with the apical O vacancies, shows that the rocksalt cation arrangement exhibits a much larger shift current susceptibility, albeit with a higher onset photon energies, whereas its Glass coefficient is smaller at lower energies (\textless2.8 eV) but greater at higher energies (\textgreater2.8 eV) with respect to that of the layered arrangement (Fig.~\ref{SC-2unit}).
Simultaneously, the imaginary dielectric constant, which is proportional to the transition rate, of the rocksalt cation arrangement is also smaller for photon energies smaller than 2.8 eV, but greater for the photon energies above 2.8 eV.
Comparison of the shift vector shows that even though the rocksalt arrangement has a smaller shift vector for the whole spectral range, the difference in the magnitude is decreasing with increasing photon energies.
Figure~\ref{PDOS-112-RS} shows the PDOSs of both the $1\times1\times2$ layered and rocksalt cation arrangements.
Clearly, the top of the VB has Ni 3$d_{zy/zx}$ orbitals in the layered arrangement, but the VB has Ni 3$d_{xy}$ orbitals in the rocksalt arrangement, leading to a preferred motion of the shift current carriers and a larger onset Glass coefficient in the layered arrangement for the electronic transitions at the band edge.
However, as the absorbed photon energies increase and states below the VB becomes involved, Ni 3$d_{z^{2}}$ orbitals become more important for the rocksalt cation arrangement while Ni 3$d_{xy}$ orbitals play a greater role for the layered arrangement. This leads to a steady enhancement of the shift vector magnitude in the rocksalt but not in the layered cation arrangement.

Furthermore, the electronic states of both the VBs and CBs in the rocksalt cation arrangement are much more localized than those in the layered structure.
This gives rise to a more sharply peaked DOS contributing to electronic transitions in a narrower range of photon energies [Fig.~\ref{PDOS-112-RS}(c)], in agreement with the overall greater transition rate in the rocksalt cation arrangement.
Combined, these two effects dictate the greater shift current susceptibility in the rocksalt cation arrangement.
These features are not readily evident from the Glass coefficient, as the Glass coefficient includes the light attenuation effect represented by the absorption coefficient (imaginary dielectric constant) that relates to the strength of the electronic transitions.
Consequently, the Glass coefficient of the rocksalt cation arrangement is only moderately larger than that of the layered cation arrangement and only for high photon energies.
This more localized nature of the electronic states in the rocksalt cation arrangement can be ascribed to its structural properties.
Compared to the layered arrangement, in the rocksalt cation arrangement the -Ni-V$_{\rm O}$-Ni- network is interrupted by the Nb atoms, leading to more localized Ni 3$d$ orbital states in the CB.
This narrower bandwidth of the CB not only induces a larger band gap (Table~\ref{tab:HSE}) and a higher shift current onset photon energy, but also an overall enhancement of the shift current magnitude.
Therefore, we see that the resulting shift current is significantly affected by the orbital character of the electronic transitions and the localization of these electronic states, which are in turn affected by the structural properties including the lattice asymmetry, the cation arrangement, and the location of the O vacancies.  
\subsection{The shift current of KBNNO with a lower concentration of O vacancies}
\begin{figure}[t]
\centering
\subfigure{\includegraphics[width=0.6\textwidth]{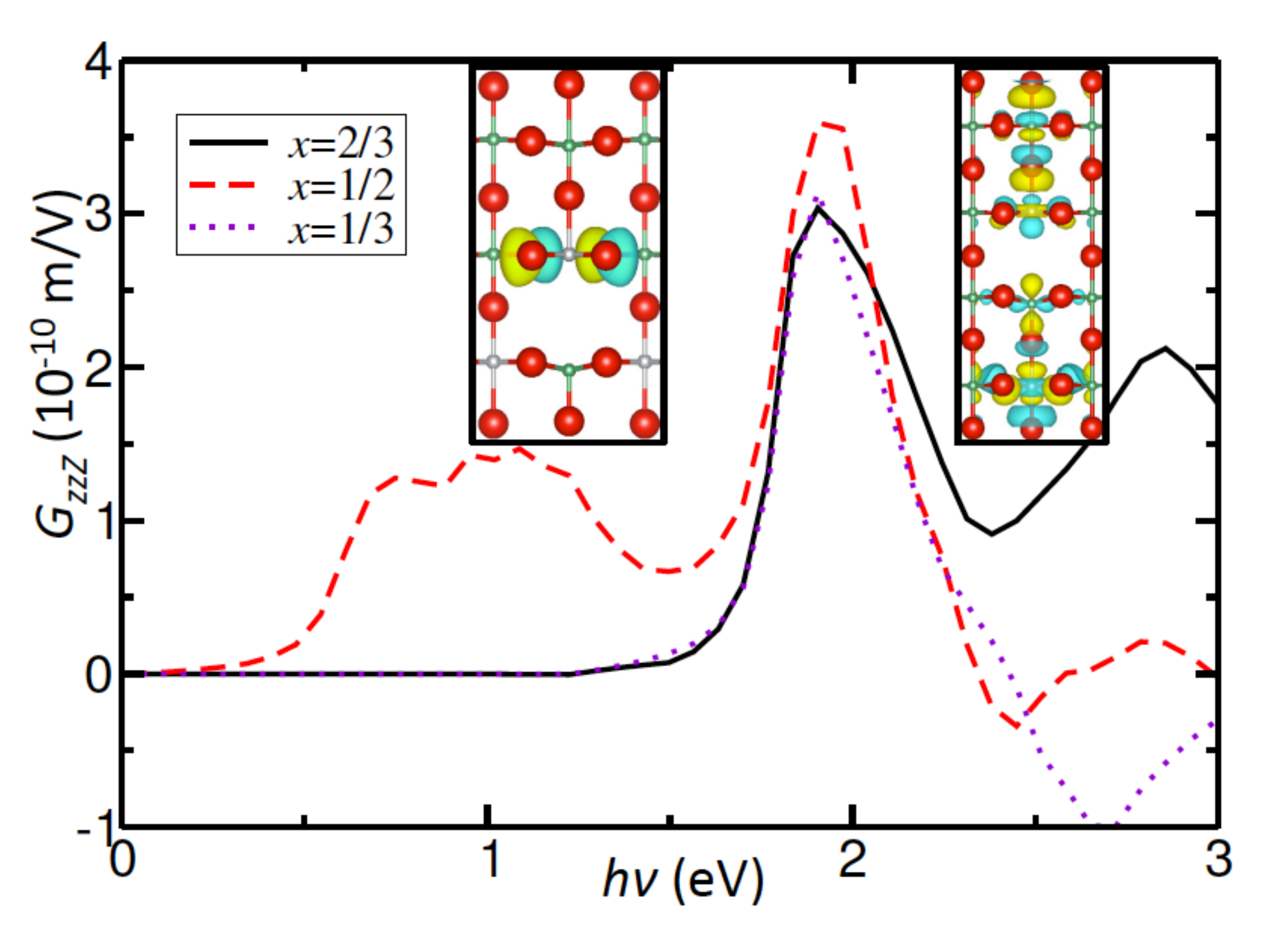}}
\caption{(Color online) The Glass coefficient of the (1-$x$)KNbO$_{3}$-$x$Ba(Ni$_{1/2}$Nb$_{1/2}$)O$_{11/4}$ solid solution with different compositions ($x$=2/3, 1/2, and 1/3). The left and right insets show the real-space wavefunction of the valence band maximum (VBM) of the 1/3KNbO$_{3}$-2/3Ba(Ni$_{1/2}$Nb$_{1/2}$)O$_{11/4}$ ($\sqrt2\times\sqrt2\times3$) and 1/2KNbO$_{3}$-1/2Ba(Ni$_{1/2}$Nb$_{1/2}$)O$_{11/4}$ ($\sqrt2\times\sqrt2\times4$) solid solutions, respectively. The wavefunction in the right inset has an extended nature along the Cartesian $z$ direction, while that in the left inset is only distributed within the $xy$ plane.
\label{G-composition}}
\end{figure}
%
The above solid solutions have a fairly high concentration of O vacancies, which could impede the motion of the photocurrent carriers, because they may behave as recombination centers.
If so, it would be necessary to reduce the amount of O vacancies while preserving the beneficial effects of O vacancies in reducing the band gap and enhancing the visible-light absorption.
In the following, we study the shift current of KBNNO with a lower concentration of O vacancies.
These KBNNO solid solutions have the compositions of (1-$x$)KNbO$_{3}$-$x$Ba(Ni$_{1/2}$Nb$_{1/2}$)O$_{11/4}$ ($x$=2/3, 1/2, and 1/3), with a vacancy concentration of 5.6\%, 4.2\%, and 2.7\%, respectively.
The corresponding supercells are $\sqrt2\times\sqrt2\times3$, $\sqrt2\times\sqrt2\times4$, and 2$\times2\times3$, respectively.
In each supercell, two Nb$^{5+}$ ions are replaced with two Ni$^{2+}$ ions, and the charge is compensated by the combination of an O vacancy adjacent to Ni and four K$^{+}$
 ions randomly substituted by the Ba$^{2+}$ ions (Ba$^{\dotr}_{\rm K}$).

Fig.~\ref{G-composition} shows the calculated Glass coefficient of different KBNNO compositions.
All three solid solutions exhibit a maximum Glass coefficient with photon energies $\approx$1.9 eV.
Furthermore, the $x$=1/2 composition has the lowest onset photon energy, as its direct band gap is the smallest among all three KBNNO solid solutions. 
It also exhibits the largest Glass coefficient (3.8$\times$10$^{-10}$ m/V) for the photon energies below 3.0 eV, which is ten times larger than the maximum Glass coefficient of BiFeO$_{3}$ (0.4$\times$10$^{-10}$ m/V).
Similarly, the difference in the magnitude of the Glass coefficient is attributed to the different orbital compositions of the contributing electronic states.
For the $x$=2/3 composition, the top of the VB is predominantly composed of O 2$p$ orbitals combined with a slight contribution of the Ni and Nb $d_{zy/zx}$ orbitals.
However, there is an extensive contribution of the Ni and Nb $d_{z^{2}}$ orbitals to the top of the VB for $x$=1/2 (Fig.~\ref{G-composition}).
This more extended wavefunction nature along the shift current direction allows for an easier motion of the shift current carriers, corresponding to the larger Glass coefficient for $x$=1/2.

%
\begin{figure}[t]
\centering
\includegraphics[width=0.6\textwidth]{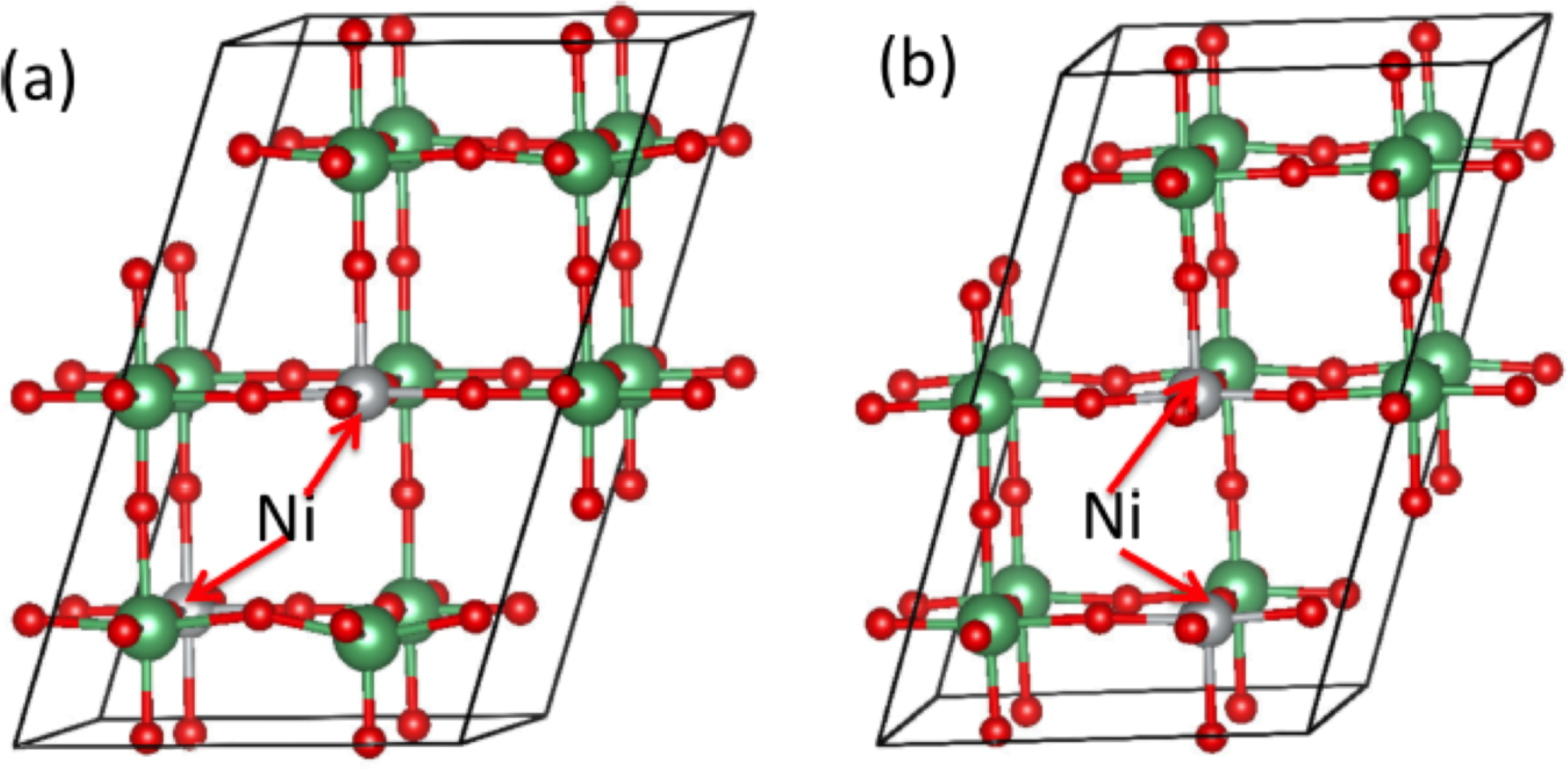}
\subfigure{\includegraphics[width=0.45\textwidth]{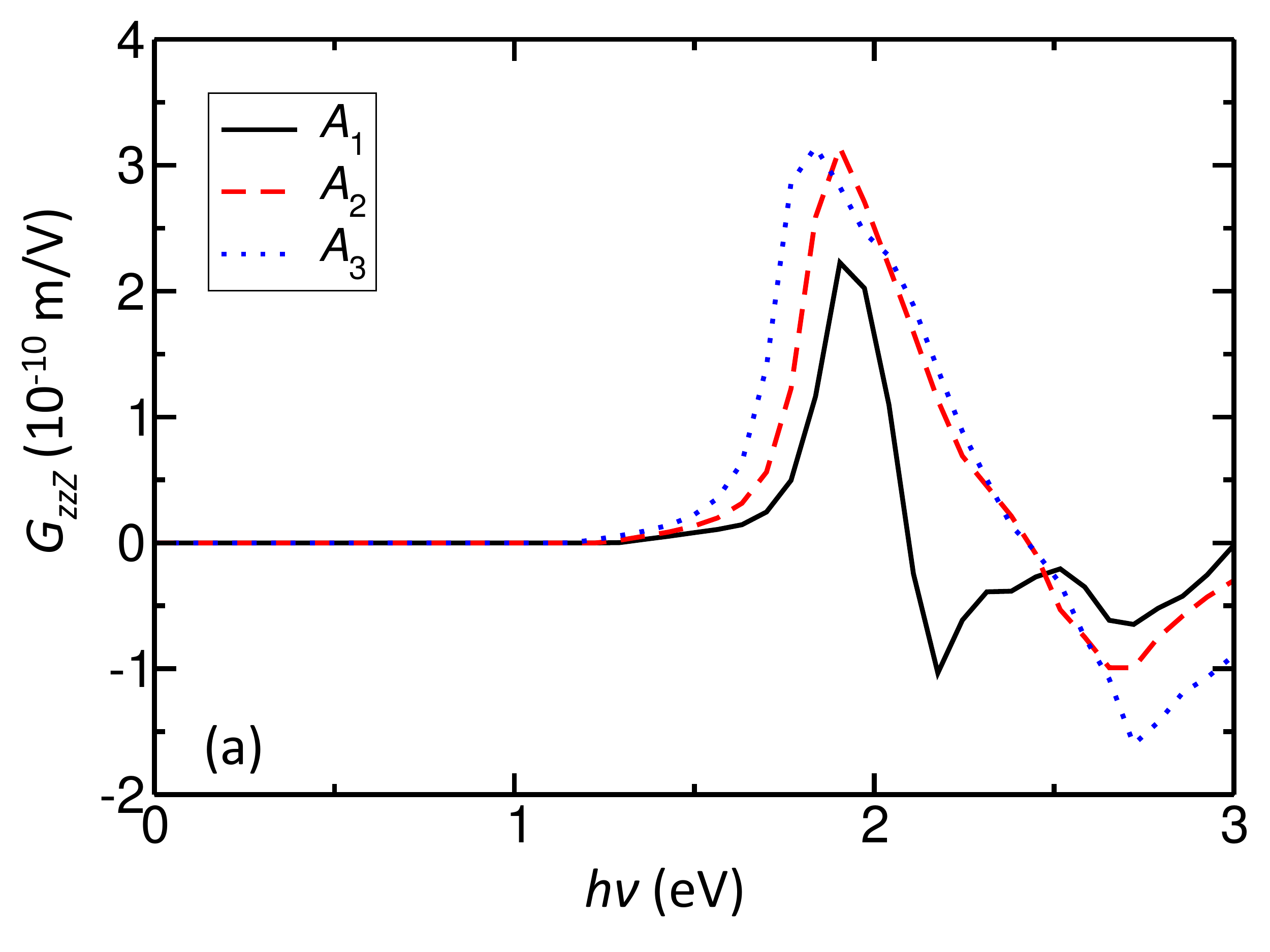}}
\subfigure{\includegraphics[width=0.45\textwidth]{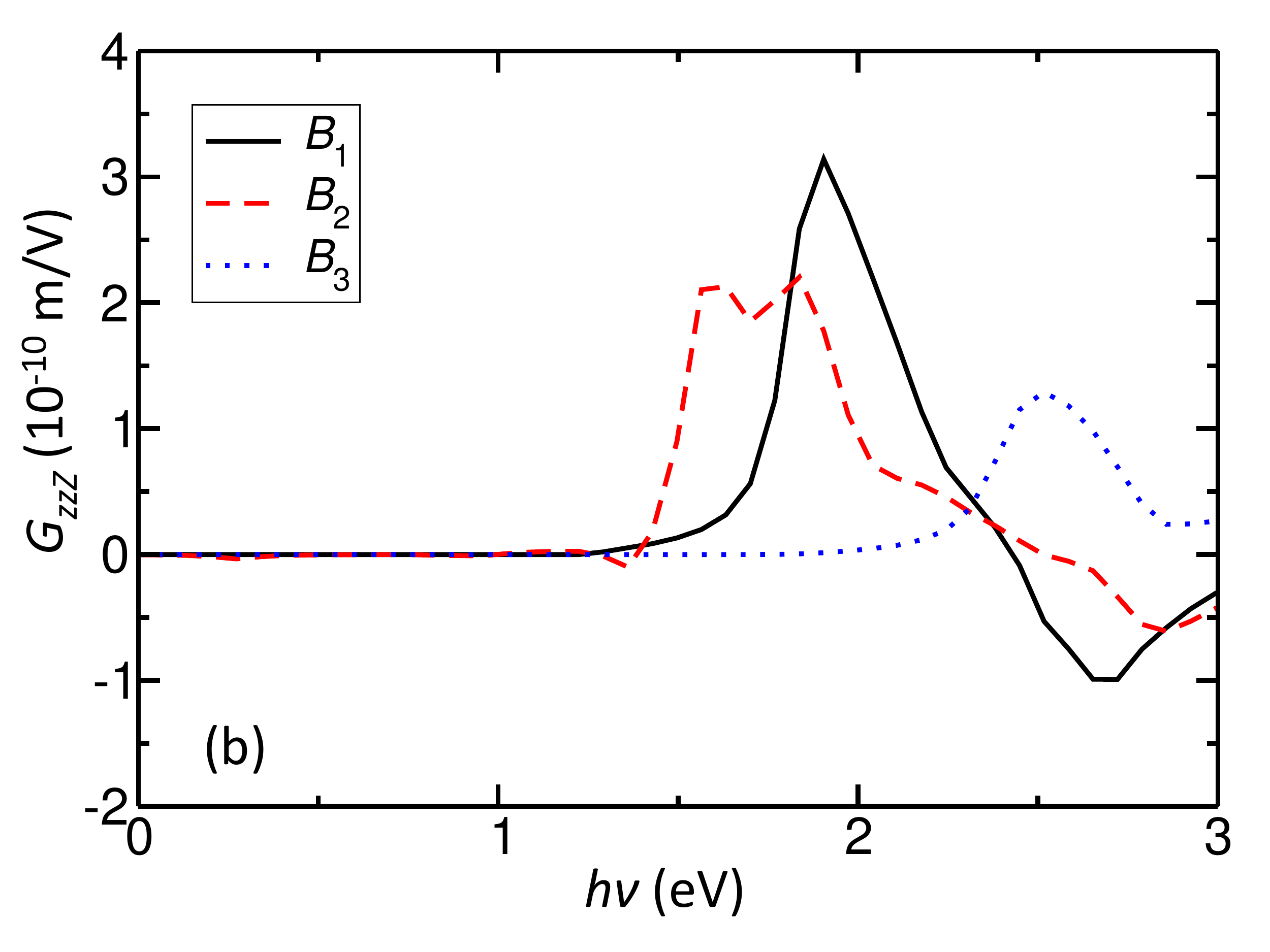}}
\caption{(Color online) The atomic structure representation of the 2/3KNbO$_{3}$-1/3Ba(Ni$_{1/2}$Nb$_{1/2}$)O$_{11/4}$ ($2\times2\times3$) solid solution with (a) two Ni$^{2+}$ ions at the body diagonal positions ($B_{1}$) (b) two Ni$^{2+}$ ions along the Cartesian $z$ direction ($B_{3}$). The atoms are shown with their accepted atomic radii. The K$^{+}$ and Ba$^{2+}$ are omitted for clarity.
The Glass coefficients of the 2/3KNbO$_{3}-$1/3Ba(Ni$_{1/2}$Nb$_{1/2}$)O$_{11/4}$ solid solutions with different (c) $A$-site and (d) $B$-site cation arrangements. The $B$-cation arrangement remains the same (with two Ni$^{2+}$ ions distributed along the body diagonal direction) when the $A$-cation arrangement is varied in (c), whereas the $A$-cation arrangement remains the same when the $B$-cation arrangement is varied in (d). ``$A_{1}$", ``$A_{2}$", and ``$A_{3}$" correspond to configurations with the four Ba$^{2+}$ cations distributed within 1, 2, and 3 layers, while ``$B_{1}$", ``$B_{2}$", and ``$B_{3}$" correspond to the configurations that the two Ni$^{2+}$ cations are distributed along the body-diagonal, face-diagonal, and the Cartesian $z$ directions. ``$A_{2}$" and ``$B_{1}$" are the same structure.
\label{12KNO-Struc}}
\end{figure}
Previous experiment has shown that the $x$=0.1 KBNNO solid solution has the lowest band gap and exhibits the best photovoltaic performance~\cite{Grinberg13p509}. Therefore, in the following we choose the $2\times2\times3$ supercell that gives our lowest calculated vacancy concentration of 2.7\% ($x$=1/3) to study the effect of different $A$- and $B$-cation arrangements on shift current.
There are two Ni$^{2+}$ ions, four Ba$^{2+}$ ions and one O vacancy in each supercell.
First, the shift current is calculated with the four Ba$^{2+}$ cations distributed over 1, 2, and 3 different layers (indicated as $A_{1}$, $A_{2}$, and $A_{3}$) while the two Ni$^{2+}$ cations are kept at the body diagonal positions with respect to each other~(Fig.~\ref{12KNO-Struc}).
The $A$ cation arrangement only has a slight effect on shift current, including both its magnitude and photon energies.
This is because the ionic radii of the K$^{+}$ (1.64 \AA) and Ba$^{2+}$ (1.61 \AA) ions are quite similar, and thereby the change of the distribution of the Ba$^{2+}$ ions has only a minor effect on the overall structure.
Also, the valence state of the $A$ cation has a very delocalized $s$ orbital character, which also has a minor effect on the electronic structure.
As a result, all three structures with different $A$-cation arrangements have nearly identical electronic properties, with an impurity state above the top of the VB arising mainly from the O 2$p$ and Ni 3$d$ orbitals.
However, when the $B$-cation arrangement is varied, there is a substantial difference in both the Glass coefficient magnitude and the photon energies that induce the largest Glass coefficient.
Specifically, the $B_{3}$ cation arrangement has the lowest Glass coefficient, with the highest onset photon energy.
The two Ni$^{2+}$ ions in this arrangement are aligned along the Cartesian $z$ direction, with an O vacancy in between them, forming a Ni-V$_{\rm O}$-Ni complex.
Moreover, the $B_{1}$ cation arrangement with two Ni$^{2+}$ ions distributed along the body diagonal direction  exhibits the largest Glass coefficient with a moderately large onset photon energy, whereas the Glass coefficient of the $B_{2}$ cation arrangement is the second largest, but its onset photon energy is the lowest.

%
\begin{figure}[t]
\centering
\includegraphics[width=\textwidth]{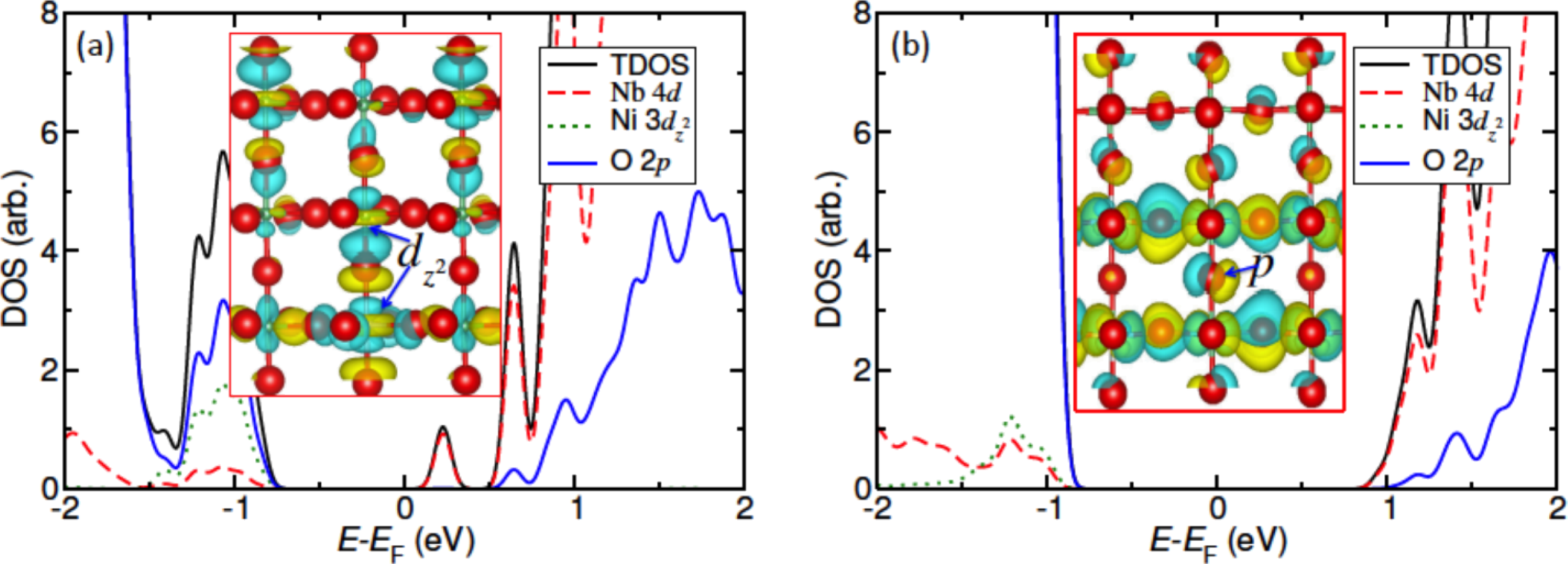}
\caption{(Color online) The densities of states (DOSs) of the 2/3KNbO$_{3}-$1/3Ba(Ni$_{1/2}$Nb$_{1/2}$)O$_{11/4}$ solid solutions with (a) the two Ni$^{2+}$ cations aligned along the body-diagonal direction ($B_{1}$) and (b) the two Ni$^{2+}$ cations aligned along the Cartesian $z$ direction ($B_{3}$). The inset shows the real-space wavefunction distribution for the top of the VB state, which shows a great contribution of Ni 3$d_{z^{2}}$ (also some Nb 4$d_{z^{2}}$) orbital character.
\label{DOS-AB}}
\end{figure}
\begin{figure}[t]
\centering
\subfigure{\includegraphics[width=0.6\textwidth]{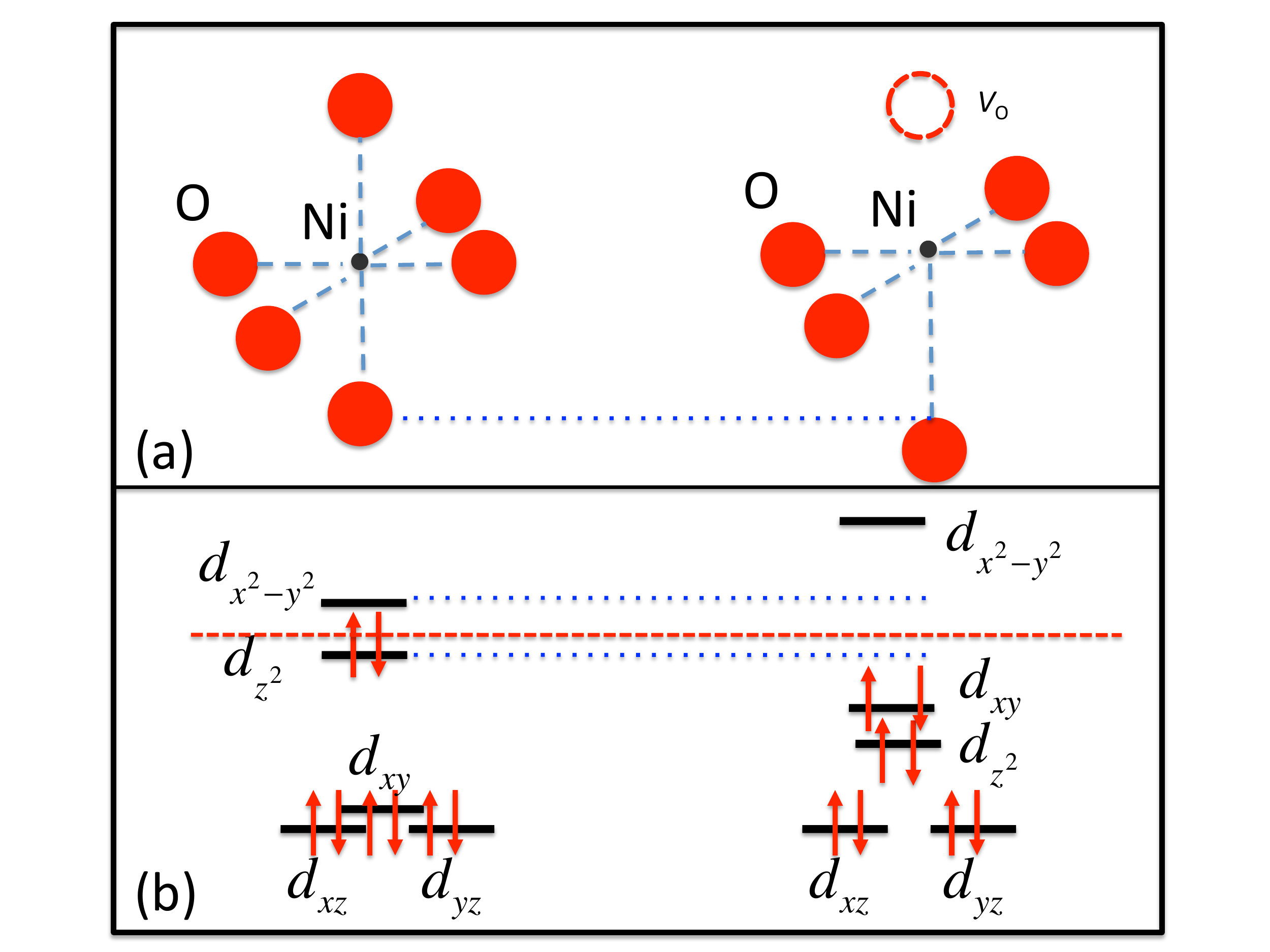}}
\caption{(Color online) The schematic representation of crystal field splitting of the Ni$^{2+}$ 3$d$ orbitals under different crystalline environments. The Ni-O distance along the Cartesian $z$ direction is longer than that in the $xy$ plane.  
\label{d-splitting}}
\end{figure}

We choose the $B_{1}$ and $B_{3}$ cation arrangements as typical examples to study the underlying origin of the difference in their shift current responses.
PDOS analysis shows that the $B_{1}$ and $B_{3}$ cation arrangements have very different electronic structure properties.
The $B_{1}$ arrangement has a gap state just above the top of the VB, leading to a smaller band gap and lower shift current onset photon energies with respect to those in the $B_{3}$ arrangement.
Furthermore, there is a predominant Nb and Ni $d_{z^{2}}$ orbital character in the $B_{1}$ arrangement, but in the $B_{3}$ arrangement the O $2p$ orbitals make the major contribution (Fig.~\ref{DOS-AB}).
The extended $d_{z^{2}}$ wavefunction nature along the $z$ direction is beneficial for the shift current response, resulting in a larger Glass coefficient in the $B_{1}$ arrangement.
This difference in the electronic properties is ascribed to their different structural properties.
Basically, there are two Ni$^{2+}$ ions and one O vacancy in each supercell.
In the $B_{1}$ arrangement, these two Ni$^{2+}$ ions have different crystal environments: one with five adjacent O atoms (NiO$_{5}$), the other with six (NiO$_{6}$).
However, in the $B_{3}$ cation arrangement, the two Ni$^{2+}$ ions share the same O vacancy, corresponding to a NiO$_{5}$ environment for both Ni$^{2+}$ ions.
For the octahedral $B$O$_{6}$ complex, the $d$ orbitals of the $B$ cation split into triply-degenerate $t_{2g}$ and doubly-degenerate $e_{g}$ states.
Because the Nb$^{5+}$ ion is strongly ferroelectric, its off-center displacement leads to a concurrent change of the Ni-O distance along the $z$ direction for the NiO$_{6}$ complex.
This induces additional splittings between the $d_{z^{2}}$ and $d_{x^{2}-y^{2}}$ orbitals of the $e_{g}$ state and between the $d_{xy}$ and $d_{zy/zx}$ orbitals of the $t_{2g}$ state~(Fig.~\ref{d-splitting}).
This change of the structural asymmetry is only moderate.
However, for the NiO$_{5}$ complex, the removal of one O atom at the apical site provides space for the O atom at the opposite site to move away from the central Ni atom.
Correspondingly, the Ni-O distance along the Cartesian $z$ direction is much larger than that in the $xy$ plane, giving rise to a near-square-planar symmetry, as shown in Fig.~\ref{d-splitting}.
Compared to the distorted NiO$_{6}$ environment, the splitting between the $d_{z^{2}}$ and $d_{x^{2}-y^{2}}$ orbitals is much larger for the NiO$_{5}$ complex.
Therefore, the energy of the $d_{z^{2}}$ orbital in the NiO$_{5}$ complex is much lower than that in the NiO$_{6}$ complex.
As a result, the higher energy of the $d_{z^{2}}$ orbital in the NiO$_{6}$ complex induces a gap state
for the $B_{1}$ cation arrangement.
The $d_{z^{2}}$ nature of the gap state generates a larger shift current response, as indicated by the Glass coefficient (Fig.~\ref{12KNO-Struc}).

The calculated maximum shift current susceptibility of the 2$\times$2$\times$3 solid solution is 1.5$\times$10$^{-4}$ Acm$^{-2}$/Wcm$^{-2}$, which is six times as large as the experimental observation for KBNNO (the samples are not completely poled)~\cite{Grinberg13p509}, and is also comparable to that of BiFeO$_{3}$.
Experiment has found that the short-circuit photocurrent of KBNNO is  0.1 $\mu$A/cm$^{2}$ under 4 mW/cm$^{2}$ illumination with above-band-gap light, corresponding to a current density of 0.25$\times$10$^{-4}$ Acm$^{-2}$/Wcm$^{-2}$, while for BiFeO$_{3}$, the current yield is 4 $\mu$A/cm$^{2}$ under 10 mW/cm$^{2}$ illumination with green light, for a current density of 4$\times$10$^{-4}$ Acm$^{-2}$/Wcm$^{-2}$~\cite{Grinberg13p509, Choi09p63}.
If we include the light attenuation effect in a thick sample of KBNNO, the photocurrent evolution is 0.4 mA/cm$^{2}$ for a  100-nm-thick sample under the illumination of a 1000 W/m$^{2}$ solar simulator, as estimated by $J_{Z}$=$G_{zzZ}\times I_{0}/d$, where $d$ and $I_{0}$ are the sample thickness and light intensity.

\subsection{The effect of strain on shift current}
\begin{figure}[t]
\centering
\includegraphics[width=0.48\textwidth]{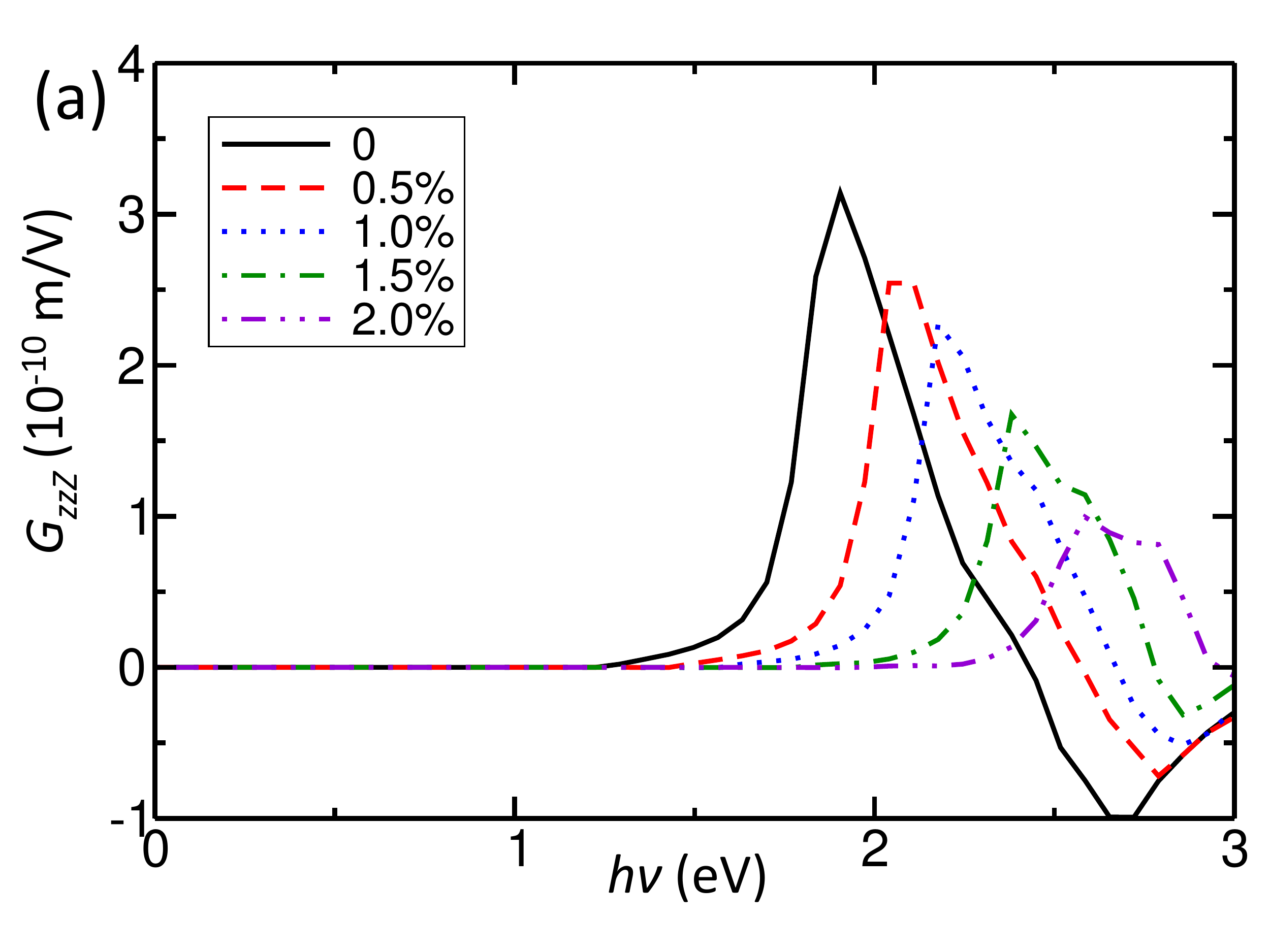}
\includegraphics[width=0.48\textwidth]{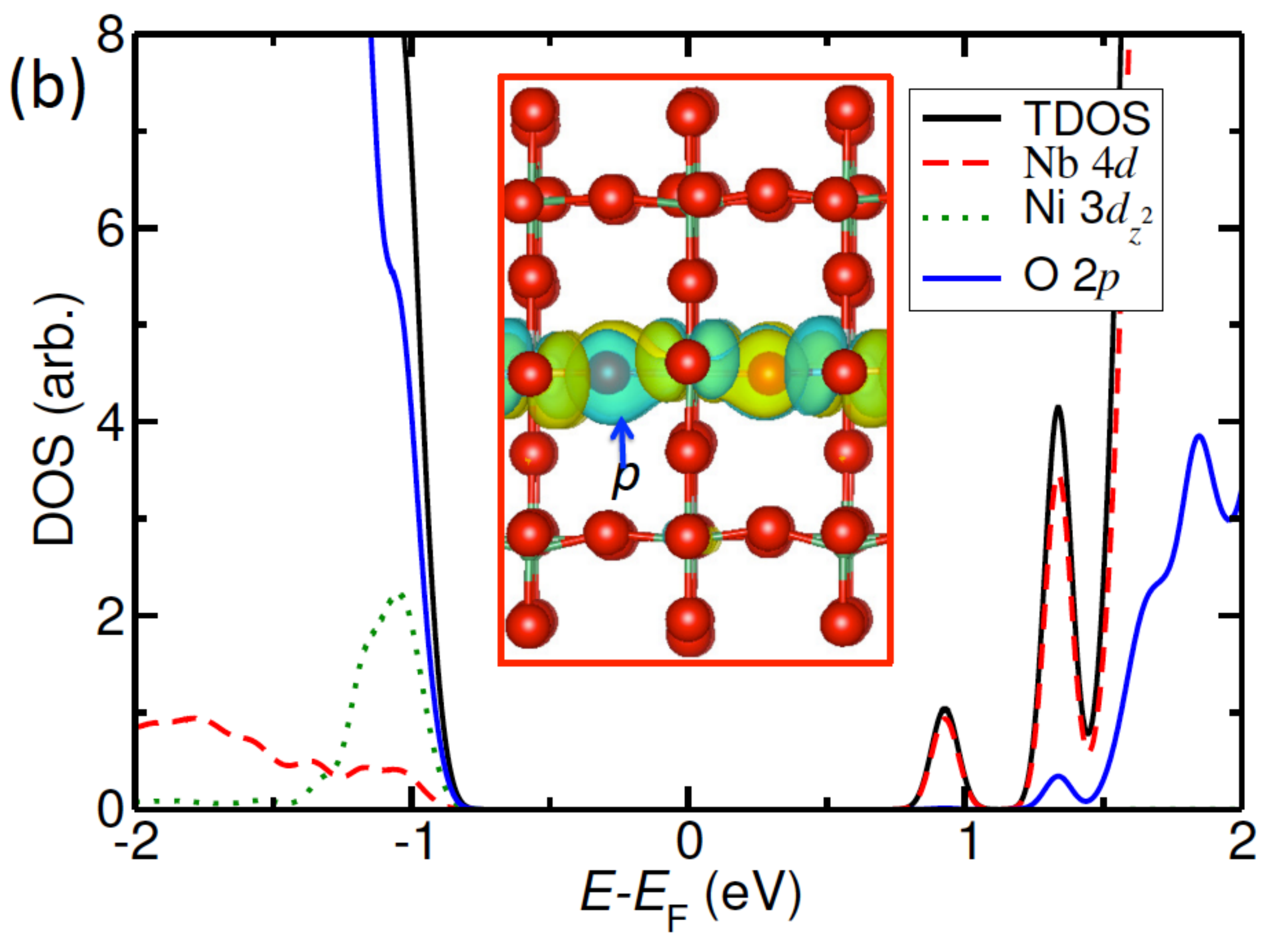}
\caption{(Color online) (a) The Glass coefficient of the 2/3KNbO$_{3}$-1/3Ba(Ni$_{1/2}$Nb$_{1/2}$)O$_{11/4}$ solid solution with different in-plane compressive strains. (b) The projected density of states (PDOS) of 2/3KNbO$_{3}$-1/3Ba(Ni$_{1/2}$Nb$_{1/2}$)O$_{11/4}$ solid solution under 2\% in-plane compressive strain. The inset in (b) shows the real-space wavefuction distribution (side view) for the top of the VB state. The solid solution is with the ``$B_{1}$" arrangement; the two Ni atoms are distributed along the body diagonal positions and the O vacancy is at the apical site of one Ni atom.
\label{Strain}}
\end{figure}
It has been shown that strain can substantially affect the octahedral cage distortions, rotate the polarization, and change the band gap of perovskite oxides~\cite{Fu00p281,Wang14p152903, Kou11p2381}.
Each of these can have a significant effect on the material's shift current response.
In order to study the effect of strain on shift current, we apply in-plane compressive strains to the 2/3KNbO$_{3}$-1/3Ba(Ni$_{1/2}$Nb$_{1/2}$)O$_{11/4}$ solid solution, in which the two Ni atoms are distributed along the body diagonal positions ($B_{1}$).
In this configuration, the O vacancy is adjacent to only one Ni atom, leaving an O$_{5}$ environment around this Ni, but an O$_{6}$ environment to the other Ni.
Figure~\ref{Strain} shows the calculated Glass coefficient as a function of the in-plane compressive strain.
It is evident that the magnitude of the Glass coefficient decreases steadily with increasing in-plane compressive strain.
Concurrently, the shift current onset photon energy also increases, suggesting a bigger band gap with enhanced strains.
This is contrary to the naive expectation that the magnitude of the Glass coefficient will increase with strain because of the polarization rotation towards the [001] direction and the overall enhancement of the structural asymmetry along the $z$ direction when applying in-plane compressive strains.

However, this change can also be rationalized by the electronic structure and wavefunction analysis.
PDOS analysis shows that the impurity state that is originally above the top of the VB shifts downwards and finally merges into it as the in-plane compressive strain increases [Figs.~\ref{DOS-AB}(a) and \ref{Strain}(b)].
Concurrently, the VB edge becomes predominantly composed of O 2$p$ orbitals [Fig.~\ref{Strain}(b)], in comparison to the significant contribution of the $d_{z^{2}}$ orbital for the configuration without strain.
This not only gives rise to a larger band gap, but also affects the motion of the shift current carriers, leading to the observed reduction of the Glass coefficient.
Further examination of the structure shows that the difference between the Ni-O distance in plane and that along the $z$ direction is substantially enhanced when the in-plane compressive strain increases.
This results in a larger splitting between the $d_{z^{2}}$ and $d_{x^{2}-y^{2}}$ orbitals for the Ni with an O$_6$ environment, becoming similar to the Ni atom in an O$_{5}$ environment (Fig.~\ref{d-splitting}).
This shifts downwards the original $d_{z^{2}}$ orbital dominated gap state, giving rise to both a larger band gap and a smaller Glass coefficient.
%
\section{CONCLUSIONS}
In summary, we study the bulk photovoltaic effect of the prototypical KNbO$_{3}$ and visible-light ferroelectric photovoltaic (K,Ba)(Ni,Nb)O$_{3-\delta}$ from first principles.
The effect of lattice distortion, oxygen vacancies, cation arrangement, composition, and strain on shift current are systematically studied.
We find that the maximum shift current response (with UV absorption) of the orthorhombic KNbO$_{3}$ is more than twice that of BiFeO$_{3}$, although KNbO$_{3}$ has a wider band gap.
Furthermore, the band-edge shift current response of the tetragonal KNbO$_{3}$ is smaller than that of its rhombohedral counterpart.
This occurs because a more isotropic lattice distortion in the rhombohedral phase reduces the splitting between the $d_{xy}$ and $d_{zx/zy}$ orbitals, leading to a larger electronic contribution of the $z$-direction-extended state to the shift current.

In (K,Ba)(Ni,Nb)O$_{3-\delta}$, the charge compensation affects significantly the favorable location of oxygen vacancies, with more stable O vacancy location based on more effective charge compensation.
The layered arrangement with apical O vacancies exhibits a much larger shift current response than that with equatorial O vacancies, as the lattice asymmetry is enhanced in the former, but reduced in the latter case.
Compared to the layered arrangement, the rocksalt arrangement has more localized electronic states, giving rise to a larger density of electronic transitions within a narrower energy range. 
Combined with the effect of the more extended $d_{z^{2}}$ orbital, this gives rise to a larger shift current susceptibility.
The effect of composition and $A$-cation arrangement on shift current is moderate, whereas the $B$-cation arrangement substantially affects both the electronic structure and the shift current, which can be rationalized by crystal field theory analysis.
With an O$_{6}$ environment around some of the Ni, the band gap is lowered, and the final shift current yield is enhanced.
The effect of strain on shift current is indirect, through the change of the wavefunction nature of the contributing electronic states, which can be used to engineer the shift current in a predictable fashion.
It is noteworthy that the shift current response of (K,Ba)(Ni,Nb)O$_{3-\delta}$ is comparable to that of BiFeO$_{3}$, but at a much lower photon energy.
More importantly, the order of magnitude enhancement of the shift current response can be gained by simply layering the (K,Ba)(Ni,Nb)O$_{5}$ solid solution, with its Glass coefficient reaching 12 times that of BiFeO$_{3}$.

Finally, we have demonstrated extensively that the shift current is dictated by the wavefuction nature of the contributing electronic orbitals, which are in turn unanimously related to their structural properties.
Therefore, we have built up a bridge between materials' structural properties and their photovoltaic performances, and provided a pathway for analyzing and elucidating the connections among these different physical properties.
The elucidated connection between materials' structure, electronic structure, and shift current is useful for the design of bulk photovoltaic materials and understanding their PV mechanisms.

\begin{acknowledgments}
F. W. was supported by the Department of Energy Office of Basic Energy Sciences, under Grant No. DE-FG02-07ER46431.
A. M. R. was supported by the Office of Naval Research. Computational support was provided by the High-Performance Computing Modernization Office of the Department of Defense and the National Energy Research Scientific Computing Center of the Department of Energy. We thank Dr. Ilya Grinberg and Dr. Steve M. Young for discussions.
\end{acknowledgments}

\end{document}